\documentclass[onecolumn,draftcls,12pt]{IEEEtran}
\usepackage{cite}

\ifCLASSINFOpdf
\else
  \usepackage[dvips]{graphicx}
\fi
\usepackage[cmex10]{amsmath}
\usepackage{cases}
\newtheorem{assumption}{Assumption}
\newtheorem{definition}{Definition}
\newtheorem{proposition}{Proposition}
\newtheorem{lemma}{Lemma}
\newtheorem{corollary}{Corollary}

\begin{document}
%
\title{Asymptotic Analysis of General Multiuser Detectors in MIMO DS-CDMA 
Channels}
%
%
%

\author{Keigo~Takeuchi, 
        Toshiyuki~Tanaka,~\IEEEmembership{Member,~IEEE}
        and~Toru~Yano
\thanks{K. Takeuchi is with the Department of Systems Science, 
Graduate School of Informatics, Kyoto University, Kyoto, 
606-8501 Japan (e-mail: takeuchi@sys.i.kyoto-u.ac.jp).}
\thanks{T. Tanaka is with the Department of Systems Science, 
Graduate School of Informatics, Kyoto University, Kyoto, 
606-8501 Japan (e-mail: tt@i.kyoto-u.ac.jp).}
\thanks{T. Yano is with the Department of Applied Physics and 
Physico-Informatics, Faculty of Science and Technology, Keio University, 
Kanagawa, 223-8522, Japan (e-mail: yano@thx.appi.keio.ac.jp).}}

\IEEEpubid{0000--0000/00\$00.00~\copyright~2007 IEEE}


\maketitle

\begin{abstract}
We analyze a MIMO DS-CDMA channel with a general multiuser detector 
including a nonlinear multiuser detector, using the replica method.  
In the many-user, limit the MIMO DS-CDMA channel with the multiuser 
detector is decoupled into a bank of single-user SIMO Gaussian 
channels if a spatial spreading scheme is employed. On the 
other hand, it is decoupled into a bank of single-user 
MIMO Gaussian channels if a spatial spreading scheme is not 
employed. The spectral efficiency of the MIMO DS-CDMA channel with 
the spatial spreading scheme is comparable with that of 
the MIMO DS-CDMA channel using an optimal space-time block code without 
the spatial spreading scheme. In the case of the 
QPSK data modulation scheme the spectral efficiency of the MIMO DS-CDMA 
channel with the MMSE detector shows {\it waterfall} behavior and is 
very close to the corresponding sum capacity when the system load 
is just below the transition point of the {\it waterfall} behavior. 
Our result implies that the performance of a multiuser detector taking  
the data modulation scheme into consideration can be far superior to 
that of linear multiuser detectors. 

\end{abstract}


%
\IEEEpeerreviewmaketitle

\section{Introduction}
\IEEEPARstart{F}{uture} wireless communications will utilize frequency  
multiplexing and spatial multiplexing to realize ultra high speed data 
communications \cite{Glisic04}. 
Joint maximum likelihood (ML) decoding with high complexity is required 
to mitigate multiple access interference and attain the 
greatest possible frequency and spatial multiplexing gains. 
On the other hand, an affordable receiver structure with lower complexity 
is desired practically. 
 
The key strategy to circumvent the complexity issue is to reduce a 
high-dimensional system to some lower-dimensional systems 
by making use of orthogonality. 
Strict orthogonality is utilized in Multicarrier (MC) schemes 
\cite{Glisic04,Bolcskei02,Visuri06}, 
orthogonal space-time block codes \cite{Jafarkhani05},  
and frequency-division multiple access (FDMA). 
Although the strict orthogonality reduces the complexity, 
it can degrade the performance, as it is known 
\cite{Gallager94} that FDMA is {\it inferior} to code-division multiple 
access (CDMA). A {\it better} trade-off between complexity 
and performance is provided by adopting statistical orthogonality.   
The direct-sequence CDMA (DS-CDMA) scheme, which is mathematically 
equivalent to multiple-input multiple-output (MIMO) systems with the 
Vertical Bell Labs Layered Space-Time (V-BLAST) architecture, decomposes  
joint decoding into multiuser detection and single-user decoding 
approximately, using the statistical orthogonality of spreading sequences. 
Although it has been reported in \cite{Verdu99} that information loss due to 
spreading is negligible in a highly loaded system,  
there exists a large gap between the spectral efficiency of 
a DS-CDMA channel with a conventional linear multiuser detector front 
end \cite{Tse99} and the corresponding sum capacity \cite{Verdu99}. 
The gap can be closed by nonlinear multiuser detection.   
The result in \cite{Tanaka02} implies that an optimal nonlinear 
multiuser detector has the capability of achieving very close performance 
to the corresponding sum capacity. 
An asymptotically optimal polynomial-time iterative detection algorithm 
on the basis of message passing has been proposed in \cite{Kabashima03}.  
Furthermore, Guo et al. \cite{Guo05} discovered the decoupling structure 
of the DS-CDMA channel under a general condition, 
due to the statistical orthogonality. 
An interesting observation is that the algorithm of 
\cite{Kabashima03} has the same decoupling structure as that of the 
optimal multiuser detector \cite{Ikehara07}.   

Recently, MIMO DS-CDMA systems, which are the application of MIMO 
technology to the DS-CDMA scheme, are being considered in anticipation 
of application to future wireless communications 
\cite{Mantravadi03,Juntti05,Ni05,Nordio062}. 
However, analysis of MIMO DS-CDMA systems is far from 
sufficient since many studies on MIMO DS-CDMA systems focus on 
conventional linear multiuser detectors \cite{Mantravadi03,Ni05,Nordio062},  
and it remains unclear how statistical orthogonality provided by 
both MIMO and DS-CDMA decomposes the MIMO DS-CDMA systems. 
Elucidation of the performance and structure of MIMO DS-CDMA systems 
with nonlinear multiuser detectors will help reveal the potential of 
MIMO DS-CDMA systems, and construct an algorithm of nonlinear multiuser 
detection. 
We thus perform an information-theoretical analysis of MIMO DS-CDMA 
systems under a general condition, including nonlinear multiuser detections.

The present paper is organized as follows. A MIMO DS-CDMA system model 
in a flat fading channel is introduced in Section~\ref{section_Model}. 
Our main results are summarized in Section~\ref{section_main} and 
the numerical evaluations are described in Section~\ref{section_numerical}. 
We discuss nonlinear multiuser detection and the extension of the results 
in Section~\ref{section_extension}. 
Section~\ref{section_conclusion} presents our conclusions.    

Throughout the present paper, $\boldsymbol{A}^{T}$ and $\boldsymbol{A}^{H}$ 
denote the transpose of $\boldsymbol{A}$ and the conjugate transpose of 
$\boldsymbol{A}$, respectively. 
$\mathcal{CN}(\boldsymbol{0},\boldsymbol{\Sigma})$ stands for  
the circularly symmetric zero-mean complex Gaussian distribution 
with the covariance matrix $\boldsymbol{\Sigma}$. 
$\boldsymbol{e}_{N}$ represents the $N$-dimensional vector 
of which the elements are all one. 
$p(\boldsymbol{y}|\boldsymbol{x},\{\boldsymbol{A}_{j}\};
\boldsymbol{\Theta})$ denotes the probability density 
function of $\boldsymbol{y}$ conditioned on $\boldsymbol{x}$ and 
$\{\boldsymbol{A}_{j}\}$ with the parameter $\boldsymbol{\Theta}$. 
$\log a$ and $\ln a$ stand for $\log_{2}a$ and $\log_{\mathrm{e}}a$, 
respectively. $\mathrm{KL}(\cdot ||\cdot )$ represents the 
Kullback-Leibler divergence with the logarithm to base $2$. 
$|\mathcal{A}|$, $\otimes$, $\mathrm{Re}(a)$, and $\mathrm{Im}(a)$ denotes 
the number of elements of a set $\mathcal{A}$, the Kronecker product 
operator between two matrices, the real and imaginary parts of 
a complex number $a$, respectively.

 

\section{Model} \label{section_Model}
\subsection{MIMO DS-CDMA Channel}
We consider an uplink of a chip-synchronous MIMO DS-CDMA channel with $K$ 
users \cite{Mantravadi03}. 
The $k$th user and the receiver have $M_{k}$ transmit antennas and 
$N$ receive antennas, respectively. A space-time encoder for the $k$th 
user maps its message into a codeword matrix, each element of which is 
mapped into a symbol by a data modulator. The product of 
each symbol $x_{m}^{k}$, the variance of which is $P_{m}^{k}$ 
(received power), and a random spreading sequence 
$\{s_{lm}^{k}; l=1,\ldots, L\}$ with the spreading factor $L$ is 
then transmitted from the $m$th transmit antenna of the 
$k$th user. The $N$-dimensional received signal vector 
$\boldsymbol{y}_{l}$ propagated through a flat fading channel is given by 
\begin{equation} \label{MIMO_CDMA_channel}
\boldsymbol{y}_{l} = 
\sum_{k=1}^{K}\boldsymbol{H}^{k}\boldsymbol{S}_{l}^{k}\boldsymbol{x}^{k} 
+ \boldsymbol{n}_{l}, 
\quad l=1,\ldots,L, 
\end{equation}
where $\boldsymbol{x}^{k}=(x_{1}^{k}, \ldots, x_{M_{k}}^{k})^{T}$ is 
the symbol vector for the $k$th user; 
the $M_{k}\times M_{k}$ diagonal matrix $\boldsymbol{S}_{l}^{k}$ is given by 
$\boldsymbol{S}_{l}^{k}=\mathrm{diag}(s_{l1}^{k}, \ldots, s_{lM_{k}}^{k})$; 
$\boldsymbol{H}^{k}=(\boldsymbol{h}_{1}^{k}, 
\ldots, \boldsymbol{h}_{M_{k}}^{k})$ is the $N\times M_{k}$ random channel 
matrix for the $k$th user, i.e., the ($n$, $m$)-element $h_{nm}^{k}$ 
of $\boldsymbol{H}^{k}$ represents the complex channel gain from 
the $m_{k}$th transmit antenna of the $k$th user to the $n$th 
receive antenna; and $\boldsymbol{n}_{l}\sim
\mathcal{CN}(\boldsymbol{0},N_{0}\boldsymbol{I}_{N})$ is additive 
white Gaussian noise (AWGN). 
The MIMO DS-CDMA channel~(\ref{MIMO_CDMA_channel}) can be expressed 
as the vector channel of higher dimension,  
\begin{equation}
\vec{\boldsymbol{y}} = 
\boldsymbol{\mathcal{A}}\vec{\boldsymbol{x}} 
+ \vec{\boldsymbol{n}}, 
\end{equation}
where $\vec{\boldsymbol{x}}=({\boldsymbol{x}^{1}}^{T}, \ldots , 
{\boldsymbol{x}^{K}}^{T})^{T}$ is the entire symbol vector; 
$\vec{\boldsymbol{y}}=
({\boldsymbol{y}_{1}}^{T}, \ldots , {\boldsymbol{y}_{L}}^{T})^{T}$ 
represents the entire received signal vector in a symbol period; 
$\vec{\boldsymbol{n}}=({\boldsymbol{n}_{1}}^{T}, \ldots , 
{\boldsymbol{n}_{L}}^{T})^{T}$ is the entire noise vector; 
and the matrix $\boldsymbol{\mathcal{A}}$ is given by
\begin{equation} \label{channel_matrix}
\boldsymbol{\mathcal{A}} = 
\begin{bmatrix}
\boldsymbol{H}^{1}\boldsymbol{S}_{1}^{1} & \cdots & 
\boldsymbol{H}^{K}\boldsymbol{S}_{1}^{K} \\
\vdots & & \vdots \\
\boldsymbol{H}^{1}\boldsymbol{S}_{L}^{1} & \cdots & 
\boldsymbol{H}^{K}\boldsymbol{S}_{L}^{K} 
\end{bmatrix}.
\end{equation} 
 
\begin{center}
 \begin{figure}[tbp]
  \includegraphics[width=\hsize]{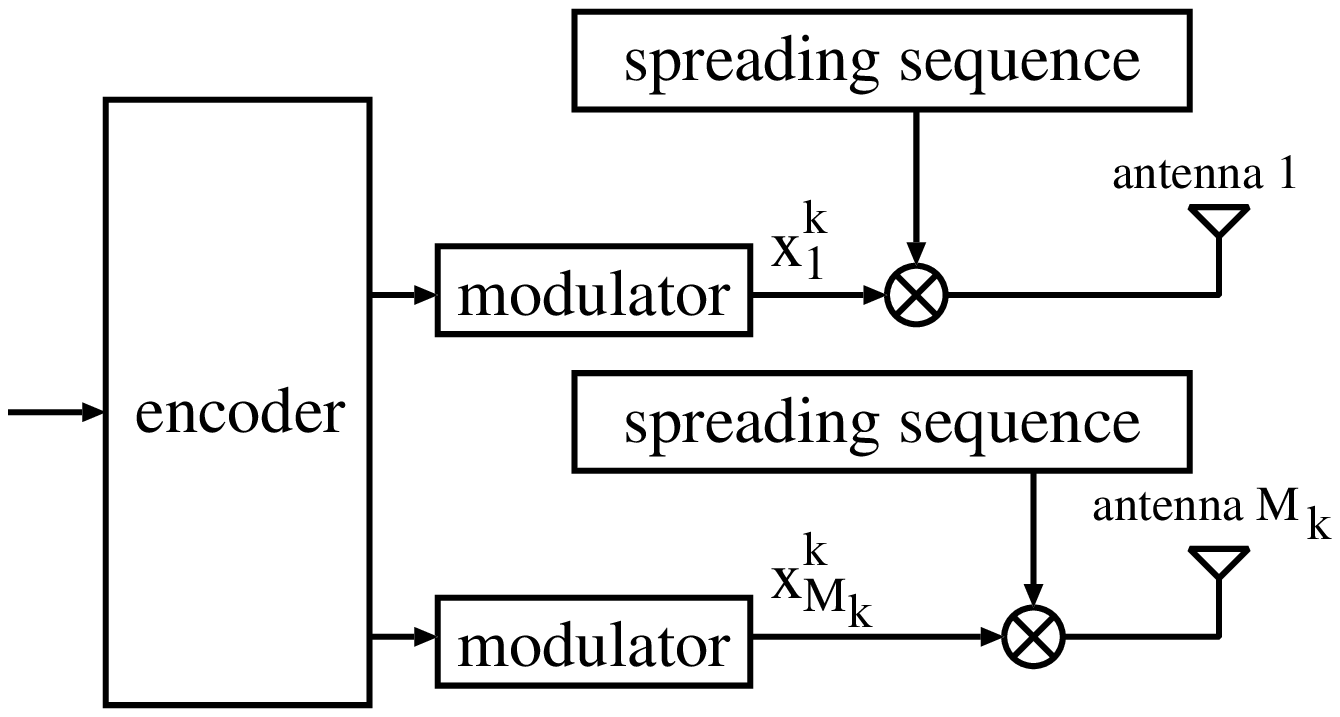}
  \caption{
  Transmitter structure with the STS scheme. 
  } 
  \label{STS_scheme}
 \end{figure}
\end{center}
\begin{center}
 \begin{figure}[tbp]
  \includegraphics[width=\hsize]{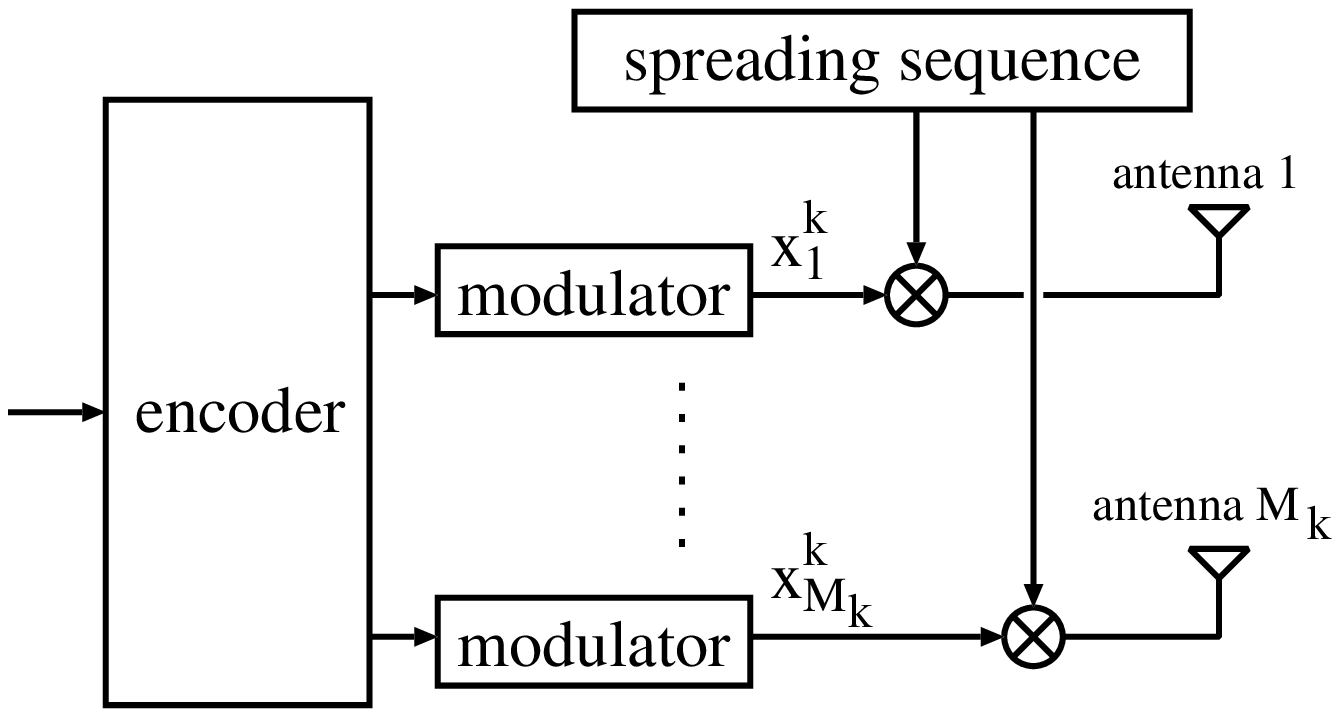}
  \caption{
  Transmitter structure with the TS scheme. 
  } 
  \label{TS_scheme}
 \end{figure}
\end{center}

We consider a space-time spreading (STS) scheme, which is implemented 
with independent spreading sequences from antenna to antenna 
(Fig.~\ref{STS_scheme}), and a time spreading (TS) scheme, which uses the 
same spreading sequences for different antennas of each user 
(Fig.~\ref{TS_scheme}). 
\begin{assumption} \label{assumption_spreading_sequence}
The real part and the imaginary part of 
$\{(s_{l1}^{k},\ldots,s_{lM_{k}}^{k})^{T};k\in\mathcal{K}=\{1,\ldots,K\},$ 
$l=1,\ldots, L\}$ 
are independent and identically distributed (i.i.d.) zero-mean random 
vectors with the covariance matrix $[\delta\boldsymbol{e}_{M_{k}}
\boldsymbol{e}_{M_{k}}^{T}+ (1-\delta)\boldsymbol{I}_{M_{k}}]/(2L)$. 
\end{assumption}

Letting $\delta=0$ and $1$ in Assumption~\ref{assumption_spreading_sequence} 
correspond to the STS and TS schemes, respectively. 
We further assume that the users are divided into $P$ groups, 
in each of which the users employ the same number of transmit 
antennas and the same data modulation scheme. 
\begin{assumption} \label{assumption_prior}
For a partition $\{\mathcal{K}_{p}; p=1, \ldots, P\}$ of $\mathcal{K}$,   
$\{\boldsymbol{x}^{k};k\in\mathcal{K}_{p}\}$ are i.i.d. complex 
random vectors, and the symbol vectors of users who belong to different 
groups are mutually independent. 
\end{assumption} 
\begin{assumption} \label{assumption_modulation}
The moment generating function of $x_{m}^{k}$ exists in the neighborhood 
of the origin. 
\end{assumption}

We remark that Assumption~\ref{assumption_prior} includes the case 
in which the number of the groups is equal to the number of users 
and that Assumption~\ref{assumption_modulation} holds in the case 
of the Gaussian data modulation as well as any conventional digital 
data modulations. Although we need these assumptions in order to derive 
part of results described in the next Section, 
we believe that Assumption~\ref{assumption_prior} can be relaxed.


\subsection{Receiver Structure}
Assuming the channel side information (CSI) only at the 
receiver, we consider separate decoding, where a multiuser detector 
front end feeds soft or hard decisions to single-user space-time decoders. 
The multiuser detector postulates that the prior distribution of the symbol 
vector for the $k$th user is $p(\tilde{\boldsymbol{x}}^{k})$, 
the moment generating function of which is assumed to exist in the 
neighborhood of the origin, and that 
the variance of the noise is $\tilde{N}_{0}$. 
Tildes indicate that the postulated 
prior distribution and the postulated variance of the noise need not 
coincide with the actual ones.       
The multiuser detector estimates the symbol vector of each user 
on the basis of the generalized posterior mean estimator (GPME) \cite{Guo05}  
\begin{equation} \label{posterior_mean_estimator}
\langle \tilde{\boldsymbol{x}}^{k} \rangle =
\frac{
 \mathrm{E}_{\vec{\tilde{\boldsymbol{x}}}}\left[
  \tilde{\boldsymbol{x}}^{k}
  p(\vec{\tilde{\boldsymbol{y}}}=\vec{\boldsymbol{y}}|
  \vec{\tilde{\boldsymbol{x}}}, \boldsymbol{\mathcal A};\tilde{N}_{0})
 \right]
}
{
 \mathrm{E}_{\vec{\tilde{\boldsymbol{x}}}}
 \left[
  p(\vec{\tilde{\boldsymbol{y}}}=\vec{\boldsymbol{y}}|
  \vec{\tilde{\boldsymbol{x}}}, \boldsymbol{\mathcal A};\tilde{N}_{0})
 \right]
}, 
\end{equation}
where $\vec{\tilde{\boldsymbol{x}}}=((\tilde{\boldsymbol{x}}^{1})^{T}, 
\ldots , (\tilde{\boldsymbol{x}}^{K})^{T})^{T}$ and  
$p(\vec{\tilde{\boldsymbol{y}}}|\vec{\tilde{\boldsymbol{x}}}, 
\boldsymbol{\mathcal A};\tilde{N}_{0})$ represents the 
channel postulated by the receiver:  
\begin{equation}
\vec{\tilde{\boldsymbol{y}}} = 
\boldsymbol{\mathcal{A}}\vec{\tilde{\boldsymbol{x}}} 
+ \vec{\tilde{\boldsymbol{n}}}, 
\quad \vec{\tilde{\boldsymbol{n}}}\sim
\mathcal{CN}(\boldsymbol{0},\tilde{N}_{0}\boldsymbol{I}_{NL}).  
\end{equation}
The multiuser detector~(\ref{posterior_mean_estimator}) corresponds to 
the minimum mean-squared error (MMSE) detector if 
$p(\tilde{\boldsymbol{x}}^{k})$ 
and $\tilde{N}_{0}$ coincide with $p(\boldsymbol{x}^{k})$ and $N_{0}$. 
The multiuser detector becomes a linear detector if 
$\tilde{\boldsymbol{x}}^{k}$ follows a zero-mean circularly symmetric 
complex Gaussian distribution. Furthermore, the linear detector 
corresponds to the linear MMSE detector 
if $\tilde{N}_{0}$ coincides with $N_{0}$. 

The maximum spectral efficiency\footnote{Note that in the present paper 
the term ``maximum spectral efficiency'' refers to the spectral efficiency 
of a channel with an optimal receiver structure on condition that a data 
modulation scheme and power allocation are specified.} per chip of the 
MIMO DS-CDMA channel with CSI at the receiver is given by the mutual 
information between 
$\vec{\boldsymbol{x}}$ and $\vec{\boldsymbol{y}}$ \cite{Tse05}  
\begin{equation}  \label{mutual_information}
\mathcal{C}_{\mathrm{joint}} = 
\frac{1}{L}I(\vec{\boldsymbol{x}};\vec{\boldsymbol{y}}) = 
\frac{1}{L}\mathrm{E}\left[
 \left.  
  \log\frac{
   p(\vec{\boldsymbol{y}}|\vec{\boldsymbol{x}}, 
   \boldsymbol{\mathcal A};N_{0})
  }
  {
   \mathrm{E}_{\vec{\boldsymbol{x}}}[p(\vec{\boldsymbol{y}}| 
   \vec{\boldsymbol{x}},\boldsymbol{\mathcal A};N_{0})]
  }
 \right| \boldsymbol{\mathcal{A}} 
\right]. 
\end{equation} 
On the other hand, the spectral efficiency per chip 
of the MIMO DS-CDMA channel with the GPME detector front end is given by 
\begin{equation} \label{mutual_information_separate}
\mathcal{C}_{\mathrm{sep}} = 
\frac{1}{L}\sum_{k=1}^{K}
I(\boldsymbol{x}^{k};\langle \tilde{\boldsymbol{x}}^{k} \rangle) = 
\frac{1}{L}\sum_{k=1}^{K}\mathrm{E}\left[ 
 \left.
  \log\frac{
   p(\langle \tilde{\boldsymbol{x}}^{k} \rangle
   |\boldsymbol{x}^{k}, \boldsymbol{\mathcal A})
  }
  {
   \mathrm{E}_{\boldsymbol{x}^{k}}[
   p(\langle \tilde{\boldsymbol{x}}^{k} \rangle | 
   \boldsymbol{x}^{k},\boldsymbol{\mathcal A})]
  }
 \right| \boldsymbol{\mathcal{A}} 
\right]. 
\end{equation}


\section{Main Results} \label{section_main}
The main results comprise analytical formulas for  
$\mathcal{C}_{\mathrm{joint}}$ (Proposition~\ref{claim_spectral_efficiency}) 
and $\mathcal{C}_{\mathrm{sep}}$ (Proposition~\ref{claim_separate}), 
and decoupling results (Proposition~\ref{Claim_decoupling}).  
First, we list some definitions. 
\begin{definition}
A single-input multiple-output (SIMO) Gaussian channel for the $m$th 
transmit antenna of the $k$th user, a postulated SIMO Gaussian channel 
for the $m$th transmit antenna of the $k$th user, a MIMO Gaussian channel 
for the $k$th user, and a postulated MIMO Gaussian channel for the $k$th 
user are respectively defined as 
\begin{subequations}
\begin{align} 
{\boldsymbol{y}_{m}^{k} = 
\boldsymbol{h}_{m}^{k}x_{m}^{k} + \boldsymbol{n}_{m}^{k}, 
\quad \boldsymbol{n}_{m}^{k}\sim\mathcal{CN}(\boldsymbol{0},
\boldsymbol{R})}, \label{SIMO_channel} \\ 
\tilde{\boldsymbol{y}}_{m}^{k} = 
\boldsymbol{h}_{m}^{k}\tilde{x}_{m}^{k} 
+ \tilde{\boldsymbol{n}}_{m}^{k}, \quad 
\tilde{\boldsymbol{n}}_{m}^{k}\sim\mathcal{CN}
(\boldsymbol{0},\tilde{\boldsymbol{R}}), \label{postulated_SIMO_channel} \\ 
\boldsymbol{y}^{k} = 
\boldsymbol{H}^{k}\boldsymbol{x}^{k} + \boldsymbol{n}^{k},
\quad \boldsymbol{n}^{k}\sim\mathcal{CN}(\boldsymbol{0},
\boldsymbol{W}), \label{MIMO_channel} \\
\tilde{\boldsymbol{y}}^{k} = 
\boldsymbol{H}^{k}\tilde{\boldsymbol{x}}^{k} 
+ \tilde{\boldsymbol{n}}^{k},
\quad \tilde{\boldsymbol{n}}^{k}\sim\mathcal{CN}
(\boldsymbol{0},\tilde{\boldsymbol{W}}), \label{postulated_MIMO_channel}  
\end{align}
\end{subequations}
where $\boldsymbol{R}$, $\tilde{\boldsymbol{R}}$, 
$\boldsymbol{W}$, and $\tilde{\boldsymbol{W}}$ are $N\times N$ 
positive definite Hermitian matrices.   
\end{definition}

\begin{definition} 
The GPME of $x_{m}^{k}$ in the SIMO Gaussian channel~(\ref{SIMO_channel}) 
and the GPME of $\boldsymbol{x}^{k}$ in the MIMO Gaussian 
channel~(\ref{MIMO_channel}) are respectively given by 
\begin{subequations} 
\begin{align}
\langle 
 \tilde{x}_{m}^{k}
\rangle_{\mathrm{SIMO}} &= 
\frac{
 \mathrm{E}_{\tilde{x}_{m}^{k}}\left[ 
  \tilde{x}_{m}^{k}
  p(\tilde{\boldsymbol{y}}_{m}^{k}=\boldsymbol{y}_{m}^{k} | \tilde{x}_{m}^{k}, 
  \boldsymbol{h}_{m}^{k}; \tilde{\boldsymbol{R}})
 \right]
}
{
 \mathrm{E}_{\tilde{x}_{m}^{k}}\left[
  p(\tilde{\boldsymbol{y}}_{m}^{k}=\boldsymbol{y}_{m}^{k} | \tilde{x}_{m}^{k}, 
  \boldsymbol{h}_{m}^{k}; \tilde{\boldsymbol{R}})
 \right]
}, \label{SIMO_posterior_mean_estimator} \\
\langle 
 \tilde{\boldsymbol{x}}^{k}
\rangle_{\mathrm{MIMO}} &= 
\frac{
 \mathrm{E}_{\tilde{\boldsymbol{x}}^{k}}\left[ 
  \tilde{\boldsymbol{x}}^{k}
  p(\tilde{\boldsymbol{y}}^{k}=\boldsymbol{y}^{k}|\tilde{\boldsymbol{x}}^{k}, 
  \boldsymbol{H}^{k}; \tilde{\boldsymbol{W}})
 \right]
}
{
 \mathrm{E}_{\tilde{\boldsymbol{x}}^{k}}\left[
  p(\tilde{\boldsymbol{y}}^{k}=\boldsymbol{y}^{k}|\tilde{\boldsymbol{x}}^{k}, 
  \boldsymbol{H}^{k}; \tilde{\boldsymbol{W}})
 \right]
}. \label{MIMO_posterior_mean_estimator}
\end{align}
\end{subequations}
\end{definition}

\begin{definition}
The maximum spectral efficiencies of the SIMO Gaussian 
channel~(\ref{SIMO_channel}) and the MIMO Gaussian 
channel~(\ref{MIMO_channel}) are respectively given by   
\begin{subequations}
\begin{align}
\mathcal{C}_{\mathrm{SIMO}}^{k,m}(\boldsymbol{R}) &= 
\mathrm{E}\left\{
 \left. 
  \log\frac{
   p(
    \boldsymbol{y}_{m}^{k} |
     x_{m}^{k}, \boldsymbol{h}_{m}^{k}; \boldsymbol{R}
   )
  }
  {
   \mathrm{E}_{x_{m}^{k}}\left[
    p(
     \boldsymbol{y}_{m}^{k} |
     x_{m}^{k}, \boldsymbol{h}_{m}^{k}; \boldsymbol{R} 
    )
   \right]
  }
 \right| \boldsymbol{h}_{m}^{k}
\right\}, \\
\mathcal{C}_{\mathrm{MIMO}}^{k}(\boldsymbol{W}) &= 
\mathrm{E}\left\{ 
 \left.
  \log\frac{
   p(\boldsymbol{y}^{k} | \boldsymbol{x}^{k}, \boldsymbol{H}^{k}; 
   \boldsymbol{W})
  }
  {
   \mathrm{E}_{\boldsymbol{x}^{k}}\left[
    p( \boldsymbol{y}^{k} | \boldsymbol{x}^{k}, \boldsymbol{H}^{k}; 
    \boldsymbol{W})
   \right]
  }
 \right| \boldsymbol{H}^{k}
\right\}. \label{spectral_efficiency_MIMO} 
\end{align}
\end{subequations}
\end{definition}

\begin{definition}
The mean-squared error of the GPME~(\ref{SIMO_posterior_mean_estimator}),  
the generalized posterior variance of the SIMO Gaussian 
channel~(\ref{SIMO_channel}), the error covariance matrix of the 
GPME~(\ref{MIMO_posterior_mean_estimator}), and the generalized 
posterior covariance matrix of the MIMO Gaussian channel~(\ref{MIMO_channel}) 
are respectively defined as  
\begin{subequations}
\begin{align}
{\mathcal E}_{\mathrm{SIMO}}^{k,m}
(\boldsymbol{R},\tilde{\boldsymbol{R}}) 
&= 
\mathrm{E}\left[
 \left.
  \left|
   x_{m}^{k} - \langle \tilde{x}_{m}^{k} \rangle_{\mathrm{SIMO}} 
  \right|^{2}
 \right| \boldsymbol{h}_{m}^{k}
\right], \\
{\mathcal V}_{\mathrm{SIMO}}^{k,m}
(\boldsymbol{R},\tilde{\boldsymbol{R}}) 
&=
\mathrm{E}\left[ 
 \left.
  \left|
   \tilde{x}_{m}^{k} - \langle \tilde{x}_{m}^{k} 
  \rangle_{\mathrm{SIMO}}
  \right|^{2}
 \right| \boldsymbol{h}_{m}^{k}
\right], \label{error_covariance} \\
\boldsymbol{\mathcal E}_{\mathrm{MIMO}}^{k}
(\boldsymbol{W},\tilde{\boldsymbol{W}}) &=
\mathrm{E}\left[
 \left.
  (
   \boldsymbol{x}^{k} 
   - \langle \tilde{\boldsymbol{x}}^{k} \rangle_{\mathrm{MIMO}} 
  )
  (
   \boldsymbol{x}^{k} 
   - \langle \tilde{\boldsymbol{x}}^{k} \rangle_{\mathrm{MIMO}}
  )^{H}
 \right| \boldsymbol{H}^{k} 
\right],  \\
\boldsymbol{\mathcal V}_{\mathrm{MIMO}}^{k}
(\boldsymbol{W},\tilde{\boldsymbol{W}}) &=
\mathrm{E}\left[
 \left. 
  (
   \tilde{\boldsymbol{x}}^{k} 
   - \langle \tilde{\boldsymbol{x}}^{k} \rangle_{\mathrm{MIMO}}
  )
  (
   \tilde{\boldsymbol{x}}^{k} 
   - \langle \tilde{\boldsymbol{x}}^{k} \rangle_{\mathrm{MIMO}} 
  )^{H}
 \right| \boldsymbol{H}^{k} 
\right]. \label{posterior_covariance}
\end{align}
\end{subequations}
\end{definition}

In order to evaluate the maximum spectral 
efficiency~(\ref{mutual_information}) analytically, we consider 
the many-user limit in which both the number of users $K$ 
and the spreading factor $L$ tend toward infinity with their ratio  
$\beta=K/L$ fixed, assuming the following:
\begin{assumption} \label{assumption_self-averaging}
The spectral efficiencies~(\ref{mutual_information}) and 
(\ref{mutual_information_separate}) satisfy the 
self-averaging property with respect to $\boldsymbol{\mathcal{S}}$, i.e., 
(\ref{mutual_information}) and 
(\ref{mutual_information_separate}) converge in probability respectively 
to the expectations of (\ref{mutual_information}) and 
(\ref{mutual_information_separate}) with respect to 
${\boldsymbol{\mathcal S}}=\{\boldsymbol{S}_{l}^{k}; l=1,\ldots, 
L, k\in\mathcal{K}\}$ in the many-user limit. 
\end{assumption}

The many-user limit has been considered in many previous studies  
\cite{Tse99,Verdu99,Guo05,Mantravadi03,Nordio062,Hanly01,Tanaka02,Guo06}. 
The self-averaging property of the MIMO DS-CDMA channel has partially been 
proved under the assumption of the Gaussian data modulation, i.e., 
the self-averaging property of (\ref{mutual_information_separate}) 
has been proved for the STS scheme \cite{Hanly01,Mantravadi03}, and 
has been proved for the TS scheme for the case of $N=1$ or 
$M_{k}\rightarrow\infty$ \cite{Mantravadi03}. 
Since proving the self-averaging property for an arbitrary prior 
distribution of the symbol vector is a challenging problem, 
in this paper, as in previous studies \cite{Tanaka02,Guo05}, 
the self-averaging property is assumed .

\begin{proposition} \label{claim_spectral_efficiency}
For arbitrary realizations of $\boldsymbol{\mathcal S}$ and  
${\boldsymbol{\mathcal H}}=\{\boldsymbol{H}^{k}; k\in\mathcal{K}\}$,   
in the many-user limit the maximum spectral 
efficiencies~(\ref{mutual_information}) of the MIMO DS-CDMA channels 
with the STS and TS schemes are respectively given by 
\begin{subequations}
\begin{align}
{\mathcal C}_{\mathrm{joint}}^{\mathrm{STS}} = 
 \beta\lim_{K\rightarrow\infty}\frac{1}{K}\sum_{k=1}^{K}
 \sum_{m=1}^{M_{k}}{\mathcal C}_{\mathrm{SIMO}}^{k,m}(\boldsymbol{R})
 + \mathrm{KL}\left(
 \mathcal{CN}\left(\boldsymbol{0}, N_{0}\boldsymbol{I}_{N}\right)||
 \mathcal{CN}\left(\boldsymbol{0}, \boldsymbol{R}\right)
\right) \quad &{\rm for}\;{\rm STS}, \label{spectral_efficiency_rs_SIMO} \\
{\mathcal C}_{\mathrm{joint}}^{\mathrm{TS}} = 
\beta\lim_{K\rightarrow\infty}\frac{1}{K}\sum_{k=1}^{K}
{\mathcal C}_{\mathrm{MIMO}}^{k}(\boldsymbol{W})
+ \mathrm{KL}\left(
 \mathcal{CN}\left(\boldsymbol{0}, N_{0}\boldsymbol{I}_{N}\right)||
 \mathcal{CN}\left(\boldsymbol{0}, \boldsymbol{W}\right)
 \right) \quad &{\rm for}\;{\rm TS}, \label{spectral_efficiency_rs_MIMO}
\end{align}
\end{subequations}
where $\boldsymbol{R}$ is a solution of the fixed-point equation 
\begin{equation} \label{R_0_SIMO_equal}
\boldsymbol{R} = 
N_{0}\boldsymbol{I}_{N} 
+ \beta\lim_{K\rightarrow\infty}\frac{1}{K}\sum_{k=1}^{K}
\sum_{m=1}^{M_{k}}{\mathcal E}_{\mathrm{SIMO}}^{k,m}
(\boldsymbol{R},\boldsymbol{R})
\boldsymbol{h}_{m}^{k}(\boldsymbol{h}_{m}^{k})^{H},   
\end{equation}
and $\boldsymbol{W}$ satisfies the fixed-point equation 
\begin{equation} \label{R_0_MIMO_equal}
\boldsymbol{W} = 
N_{0}\boldsymbol{I}_{N} 
+ \beta\lim_{K\rightarrow\infty}\frac{1}{K}\sum_{k=1}^{K}
\boldsymbol{H}^{k}\boldsymbol{\mathcal E}_{\mathrm{MIMO}}^{k}
(\boldsymbol{W},\boldsymbol{W})(\boldsymbol{H}^{k})^{H}.  
\end{equation}
In the case in which there exist multiple solutions of $\boldsymbol{R}$ and 
$\boldsymbol{W}$, one should choose the solutions so as to minimize 
the maximum spectral efficiencies~(\ref{spectral_efficiency_rs_SIMO}) and 
(\ref{spectral_efficiency_rs_MIMO}), respectively. 
\end{proposition}

The derivation of Proposition~\ref{claim_spectral_efficiency} is 
summarized in Appendix~\ref{appendix_free_energy}. 
The operational meaning of the 
expressions~(\ref{spectral_efficiency_rs_SIMO}) 
and~(\ref{spectral_efficiency_rs_MIMO}) is that the first 
terms in the right-hand sides are the spectral efficiencies of the 
MIMO DS-CDMA channel with the MMSE detector front end in the cases of 
the STS scheme and the TS scheme, respectively, and the second terms 
in the right-hand sides are the separation loss, i.e., the information 
loss due to the separation of detection and decoding. 
The following lemma is needed in order to derive the propositions 
providing this interpretation. 
\begin{lemma} \label{lemma_moment}
If $\{\boldsymbol{H}^{k};k\in\mathcal{K}_{p}\}$ are i.i.d., 
in the limit $|\mathcal{K}_{1}|\rightarrow\infty, \ldots, |\mathcal{K}_{P}|
\rightarrow\infty$, and $L\rightarrow\infty$ with their ratios 
$\{\beta_{p}=|\mathcal{K}_{p}|/L;p=1,\ldots,P\}$ fixed,    
an arbitrary joint moment of the symbol vector and 
the GPME~(\ref{posterior_mean_estimator}) for a user $k\in\mathcal{K}_{p}$ 
converges to the corresponding joint moment of the symbol vector and 
its GPME in a single-user channel, i.e., 
\begin{equation}
\lim_{|\mathcal{K}_{1}|,\ldots,|\mathcal{K}_{P}|,L\rightarrow\infty}
\mathrm{E}\left[
 \prod_{m=1}^{M_{k}}\left\{  
  [\mathrm{Re}(x_{m}^{k})]^{i_{m}^{(\mathrm{r})}}
  [\mathrm{Im}({x_{m}^{k}})]^{i_{m}^{(\mathrm{i})}}
  \langle \mathrm{Re}(\tilde{x}_{m}^{k}) 
  \rangle^{j_{m}^{(\mathrm{r})}} 
  \langle \mathrm{Im}(\tilde{x}_{m}^{k})
  \rangle^{j_{m}^{(\mathrm{i})}} 
 \right\}
\right] \nonumber
\end{equation}
\begin{subnumcases}
{=}
\mathrm{E}\left[
 \prod_{m=1}^{M_{k}}\left\{ 
  [\mathrm{Re}(x_{m}^{k})]^{i_{m}^{(\mathrm{r})}}
  [\mathrm{Im}({x_{m}^{k}})]^{i_{m}^{(\mathrm{i})}}
  \left\langle 
   \mathrm{Re}(\tilde{x}_{m}^{k}) 
  \right\rangle_{\mathrm{SIMO}}^{j_{m}^{(\mathrm{r})}}
  \left\langle 
   \mathrm{Im}(\tilde{x}_{m}^{k})
  \right\rangle_{\mathrm{SIMO}}^{j_{m}^{(\mathrm{i})}}
 \right\}
\right] & for STS, \label{joint_moment_SIMO} \\ 
 \mathrm{E}\left[
  \prod_{m=1}^{M_{k}}\left\{
   [\mathrm{Re}(x_{m}^{k})]^{i_{m}^{(\mathrm{r})}}
   [\mathrm{Im}({x_{m}^{k}})]^{i_{m}^{(\mathrm{i})}}
   \left\langle 
    \mathrm{Re}(\tilde{x}_{m}^{k}) 
   \right\rangle_{\mathrm{MIMO}}^{j_{m}^{(\mathrm{r})}}
   \left\langle 
    \mathrm{Im}(\tilde{x}_{m}^{k})
   \right\rangle_{\mathrm{MIMO}}^{j_{m}^{(\mathrm{i})}}
  \right\}
 \right] & for TS, \label{joint_moment_MIMO} 
\end{subnumcases} 
where ($\boldsymbol{R}$, $\tilde{\boldsymbol{R}}$) 
is a solution of the following fixed-point equations:
\begin{subequations}
\begin{align}
\boldsymbol{R} &= 
N_{0}\boldsymbol{I}_{N} 
+ \beta\lim_{K\rightarrow\infty}\frac{1}{K}\sum_{k=1}^{K}
\sum_{m=1}^{M_{k}}{\mathcal E}_{\mathrm{SIMO}}^{k,m}
(\boldsymbol{R},\tilde{\boldsymbol{R}})
\boldsymbol{h}_{m}^{k}(\boldsymbol{h}_{m}^{k})^{H}, \label{R_0_SIMO} \\ 
\tilde{\boldsymbol{R}} &=
\tilde{N}_{0}\boldsymbol{I}_{N} 
+ \beta\lim_{K\rightarrow\infty}\frac{1}{K}\sum_{k=1}^{K}
\sum_{m=1}^{M_{k}}{\mathcal V}_{\mathrm{SIMO}}^{k,m}
(\boldsymbol{R},\tilde{\boldsymbol{R}}) 
\boldsymbol{h}_{m}^{k}(\boldsymbol{h}_{m}^{k})^{H}, \label{R_SIMO} 
\end{align}
\end{subequations}
and ($\boldsymbol{W}$, $\tilde{\boldsymbol{W}}$) 
is a solution of the following fixed-point equations:
\begin{subequations}
\begin{align}
\boldsymbol{W} &=
N_{0}\boldsymbol{I}_{N} 
+ \beta\lim_{K\rightarrow\infty}\frac{1}{K}\sum_{k=1}^{K}
\boldsymbol{H}^{k}\boldsymbol{\mathcal E}_{\mathrm{MIMO}}^{k}
(\boldsymbol{W},\tilde{\boldsymbol{W}})(\boldsymbol{H}^{k})^{H}, 
\label{R_0_MIMO} \\ 
\tilde{\boldsymbol{W}} &= 
\tilde{N}_{0}\boldsymbol{I}_{N} 
+ \beta\lim_{K\rightarrow\infty}\frac{1}{K}\sum_{k=1}^{K}
\boldsymbol{H}^{k}\boldsymbol{\mathcal V}_{\mathrm{MIMO}}^{k}
(\boldsymbol{W},\tilde{\boldsymbol{W}})(\boldsymbol{H}^{k})^{H}. 
\label{R_MIMO}
\end{align}
\end{subequations}
In the case in which there exist multiple solutions of 
($\boldsymbol{R}$, $\tilde{\boldsymbol{R}}$) and 
($\boldsymbol{W}$, $\tilde{\boldsymbol{W}}$), one should respectively  
choose the solutions that minimize the free energies 
\begin{subnumcases}
{\frac{\beta}{\ln 2}{\mathcal F} =} 
\beta\lim_{K\rightarrow\infty}\frac{1}{K}\sum_{k=1}^{K}
\sum_{m=1}^{M_{k}}\tilde{\mathcal C}_{\mathrm{SIMO}}^{k,m}
(\boldsymbol{R},\tilde{\boldsymbol{R}})
+ N\log(\pi\mathrm{e}N_{0}) 
+ F(\boldsymbol{R},\tilde{\boldsymbol{R}}) \;\; {\rm for}\;{\rm STS}, 
\label{free_energy_ind} \\
\beta\lim_{K\rightarrow\infty}\frac{1}{K}\sum_{k=1}^{K}
\tilde{\mathcal C}_{\mathrm{MIMO}}^{k}
(\boldsymbol{W},\tilde{\boldsymbol{W}})
+ N\log(\pi\mathrm{e}N_{0}) 
+ F(\boldsymbol{W},\tilde{\boldsymbol{W}}) \;\; {\rm for}\;{\rm TS}, 
\label{free_energy_same}
\end{subnumcases}
where $\tilde{\mathcal C}_{\mathrm{SIMO}}^{k,m}(\boldsymbol{R},
\tilde{\boldsymbol{R}})$ and $\tilde{\mathcal C}_{\mathrm{MIMO}}^{k}
(\boldsymbol{W},\tilde{\boldsymbol{W}})$ are given by 
\begin{subequations}
\begin{align} 
\tilde{\mathcal C}_{\mathrm{SIMO}}^{k,m}(\boldsymbol{R},
\tilde{\boldsymbol{R}}) &= 
\int p(\boldsymbol{y}_{m}^{k}, x_{m}^{k}| 
\boldsymbol{h}_{m}^{k}; \boldsymbol{R})
\log\frac{
 p(\boldsymbol{y}_{m}^{k} | x_{m}^{k}, 
 \boldsymbol{h}_{m}^{k}; \tilde{\boldsymbol{R}})
}
{
 \mathrm{E}_{\tilde{x}_{m}^{k}}\left[
  p(\tilde{\boldsymbol{y}}_{m}^{k} = \boldsymbol{y}_{m}^{k} | 
  \tilde{x}_{m}^{k}, \boldsymbol{h}_{m}^{k}; \tilde{\boldsymbol{R}})
 \right]
}dx_{m}^{k}\boldsymbol{dy}_{m}^{k}, \label{tilde_spectral_efficiency_SIMO} \\ 
\tilde{\mathcal C}_{\mathrm{MIMO}}^{k}(\boldsymbol{W},
\tilde{\boldsymbol{W}})  &= 
\int p(\boldsymbol{y}^{k},\boldsymbol{x}^{k}| \boldsymbol{H}^{k}; 
\boldsymbol{W})
\log\frac{
 p(\boldsymbol{y}^{k} | \boldsymbol{x}^{k}, \boldsymbol{H}^{k}; 
 \tilde{\boldsymbol{W}})
}
{
 \mathrm{E}_{\tilde{\boldsymbol{x}}^{k}}\left[
  p(\tilde{\boldsymbol{y}}^{k} = \boldsymbol{y}^{k} | 
  \tilde{\boldsymbol{x}}^{k}, \boldsymbol{H}^{k}; \tilde{\boldsymbol{W}})
 \right]
}\boldsymbol{dx}^{k}\boldsymbol{dy}^{k}, \label{tilde_spectral_efficiency_MIMO}
\end{align}
\end{subequations}
and the function $F(\cdot,\tilde{\cdot})$ is defined as 
\begin{eqnarray}
F(\boldsymbol{A},\tilde{\boldsymbol{A}}) = 
\mathrm{KL}\left(
 \mathcal{CN}(\boldsymbol{0}, N_{0}\boldsymbol{I}_{N})
  \left\|
   \mathcal{CN}(\boldsymbol{0}, \tilde{\boldsymbol{A}})
  \right.
 \right) 
+ \mathrm{KL}\left(
 \mathcal{CN}(\boldsymbol{0}, \boldsymbol{A})
 \left\|
  \mathcal{CN}(\boldsymbol{0}, \tilde{\boldsymbol{A}})
 \right.
 \right) \nonumber \\
+ \mathrm{KL}\left(
 \mathcal{CN}(\boldsymbol{0}, \tilde{N}_{0}\boldsymbol{I}_{N})
 \left\|
  \mathcal{CN}(\boldsymbol{0}, \tilde{\boldsymbol{A}})
 \right.
 \right) 
- \mathrm{KL}\left(
 \mathcal{CN}(\boldsymbol{0}, \tilde{N}_{0}\boldsymbol{I}_{N})
 \left\|
  \mathcal{CN}(\boldsymbol{0}, \tilde{\boldsymbol{A}}\boldsymbol{A}^{-1}
 \tilde{\boldsymbol{A}})
 \right.
 \right). \label{function_F}
\end{eqnarray}
\end{lemma}

Lemma~\ref{lemma_moment} is derived in Appendix~\ref{appendix_joint_moment}. 
From Assumption~\ref{assumption_modulation}, the moment generating 
functions of the GPMEs~(\ref{SIMO_posterior_mean_estimator}) and 
(\ref{MIMO_posterior_mean_estimator}) are expected to exist in the  
neighborhood of the origin, respectively. Then, the joint moment 
sequences~(\ref{joint_moment_SIMO}) and (\ref{joint_moment_MIMO})  
uniquely determine the respective joint distributions. Furthermore, 
assuming that all joint moments of the symbol vector and the 
GPME~(\ref{posterior_mean_estimator}) have the self-averaging property 
with respect to $\boldsymbol{\mathcal{S}}$, we obtain the following 
decoupling result. 
\begin{proposition} \label{Claim_decoupling}
If $\{x_{m}^{k},\tilde{x}_{m}^{k};m=1,\ldots,M_{k}\}$ 
are mutually independent and $\{\boldsymbol{H}^{k};k\in\mathcal{K}_{p}\}$ 
are i.i.d., in the limit $|\mathcal{K}_{1}|\rightarrow\infty, 
\ldots, |\mathcal{K}_{P}|\rightarrow\infty$, and $L\rightarrow\infty$ 
with $\{\beta_{p}\}$ fixed, then the conditional joint distribution of 
$\boldsymbol{x}^{k}$ and $\langle \boldsymbol{x}^{k} \rangle$ conditioned on 
$\boldsymbol{\mathcal{S}}$ and $\boldsymbol{\mathcal{H}}$
converges to the joint distribution of $\boldsymbol{x}^{k}$ and 
its GPME in a single-user channel, i.e., 
\begin{subnumcases}
{p(\boldsymbol{x}^{k},\langle \tilde{\boldsymbol{x}}^{k} \rangle
|\boldsymbol{\mathcal{S}},\boldsymbol{\mathcal{H}}) 
\rightarrow} 
\prod_{m=1}^{M_{k}}p(x_{m}^{k},\langle \tilde{x}_{m}^{k} 
\rangle_{\mathrm{SIMO}}|\boldsymbol{h}_{m}^{k}) & for STS, \\ 
p(\boldsymbol{x}^{k},\langle \tilde{\boldsymbol{x}}^{k} 
\rangle_{\mathrm{MIMO}}|\boldsymbol{H}^{k}) & for TS  
\end{subnumcases}
with ($\boldsymbol{R}$, $\tilde{\boldsymbol{R}}$) and 
($\boldsymbol{W}$, $\tilde{\boldsymbol{W}}$) defined in 
Lemma~\ref{lemma_moment}.  
\end{proposition}

Although in order to simplify the derivation of 
Proposition~\ref{Claim_decoupling}, we have assumed that the symbol 
vectors and the channel matrices are i.i.d. in the same group 
$\mathcal{K}_{p}$, we conjecture that this assumption is not required.   
Proposition~\ref{Claim_decoupling} implies that the decoupling principle 
\cite{Guo05} holds in the MIMO DS-CDMA channel, i.e., in the case of the 
STS scheme the MIMO DS-CDMA channel 
(\ref{MIMO_CDMA_channel}) with the GPME detector front end 
(\ref{posterior_mean_estimator}) is decoupled into a bank of 
the SIMO Gaussian channels~(\ref{SIMO_channel}) with the GPME detector 
front ends~(\ref{SIMO_posterior_mean_estimator}) in the many-user limit,  
and in the case of the TS scheme the MIMO DS-CDMA 
channel~(\ref{MIMO_CDMA_channel}) with the GPME detector front 
end~(\ref{posterior_mean_estimator}) is decoupled into a bank of the 
MIMO Gaussian channels~(\ref{MIMO_channel}) with the GPME detector 
front ends~(\ref{MIMO_posterior_mean_estimator}) in the many-user limit.   
From this decoupling result we can evaluate the spectral 
efficiency~(\ref{mutual_information_separate}) of the MIMO DS-CDMA 
channel~(\ref{MIMO_CDMA_channel}) with the GPME detector front 
end~(\ref{posterior_mean_estimator}).   
\begin{proposition} \label{claim_separate}
If $\{x_{m}^{k},\tilde{x}_{m}^{k};m=1,\ldots,M_{k}\}$ are mutually 
independent and $\{\boldsymbol{H}^{k};k\in\mathcal{K}_{p}\}$ are i.i.d., 
the asymptotic spectral efficiency of the MIMO DS-CDMA channel with the 
GPME detector front end is given by 
\begin{subequations}
\begin{align}
\mathcal{C}_{\mathrm{sep}}^{\mathrm{STS}} = 
\beta\lim_{K\rightarrow\infty}\frac{1}{K}\sum_{k=1}^{K}\sum_{m=1}^{M_{k}}
\mathcal{C}_{\mathrm{SIMO,GPME}}^{k,m}(\boldsymbol{R},\tilde{\boldsymbol{R}}) 
\quad & {\rm for}\;{\rm STS}, \label{spectral_efficiency_separate_SIMO} \\  
\mathcal{C}_{\mathrm{sep}}^{\mathrm{TS}} = 
\beta\lim_{K\rightarrow\infty}\frac{1}{K}\sum_{k=1}^{K}
\mathcal{C}_{\mathrm{MIMO,GPME}}^{k}(\boldsymbol{W},\tilde{\boldsymbol{W}})
\quad & {\rm for}\;{\rm TS}, \label{spectral_efficiency_separate_MIMO}
\end{align}
\end{subequations}
where $\mathcal{C}_{\mathrm{SIMO,GPME}}^{k,m}(\boldsymbol{R},
\tilde{\boldsymbol{R}})$ and $\mathcal{C}_{\mathrm{MIMO,GPME}}^{k}
(\boldsymbol{W},\tilde{\boldsymbol{W}})$ are the spectral efficiency of 
the SIMO Gaussian channel~(\ref{SIMO_channel}) with the GPME detector 
front end~(\ref{SIMO_posterior_mean_estimator}) and the spectral 
efficiency of the MIMO Gaussian channel~(\ref{MIMO_channel}) with 
the GPME detector front end~(\ref{MIMO_posterior_mean_estimator}), 
respectively. 
$(\boldsymbol{R},\tilde{\boldsymbol{R}})$ and 
$(\boldsymbol{W},\tilde{\boldsymbol{W}})$ are identical to those in 
Lemma~\ref{lemma_moment}.  
\end{proposition}

$\mathcal{C}_{\mathrm{sep}}^{\mathrm{STS}}$ and 
$\mathcal{C}_{\mathrm{sep}}^{\mathrm{TS}}$ coincide 
with the first terms of (\ref{spectral_efficiency_rs_SIMO}) and 
(\ref{spectral_efficiency_rs_MIMO}), respectively, if $p(\boldsymbol{x}^{k})=
p(\tilde{\boldsymbol{x}}^{k})$ and $N_{0}=\tilde{N}_{0}$, i.e., the first 
terms represent the spectral efficiencies of the MIMO DS-CDMA channel with 
the MMSE detector front end in the respective cases. 

Although we have used the replica method to derive the results presented 
so far, the following Corollary demonstrates that 
Proposition~\ref{claim_separate} is reduced to the known results 
with the Gaussian data modulation \cite{Hanly01,Mantravadi03}, 
which have been proved rigorously with random matrix theory.  
\begin{corollary}
If $\{x_{m}^{k},\tilde{x}_{m}^{k};m=1,\ldots,M_{k}=M\}$ follow i.i.d. 
$\mathcal{CN}(0,P)$ and if $\{\boldsymbol{H}^{k}; k\in\mathcal{K}\}$ 
are i.i.d. random matrices with i.i.d. entries, 
then the spectral efficiency of the MIMO DS-CDMA channel with the LMMSE 
detector front end is given by
\begin{subequations}
\begin{align} 
\mathcal{C}_{\mathrm{LMMSE}}^{\mathrm{STS}} = 
\beta M\mathrm{E}\left[
 \log\left(
  1 + \frac{P\|h_{m}^{k}\|^{2}}{N_{R}}
 \right)
\right] \quad & {\rm for}\;{\rm STS},  \\
\mathcal{C}_{\mathrm{LMMSE}}^{\mathrm{TS}} = 
\beta\mathrm{E}\left[
 \log\det\left(
  \boldsymbol{I}_{N} + \frac{P}{N_{W}}\boldsymbol{H}^{k}
  (\boldsymbol{H}^{k})^{H}
 \right)
\right] \quad & {\rm for}\;{\rm TS},
\end{align}  
\end{subequations}
where $N_{R}$ is the unique solution of the fixed-point 
equation
\begin{equation}
N_{R} = 
N_{0} + \frac{\beta}{N}\mathrm{E}\left[
 \frac{P\|\boldsymbol{h}_{m}^{k}\|^{2}}
 {1 + P\|\boldsymbol{h}_{m}\|^{2}/N_{R}}
\right], 
\end{equation}
and $N_{W}$ is the unique solution of the fixed-point 
equation
\begin{equation}
N_{W} = 
N_{0} + \frac{\beta}{N}\min(N,M)\mathrm{E}\left[
 \frac{P\lambda}{1 + P\lambda/N_{W}}
\right]  
\end{equation} 
with $\lambda$ being the random variable following the marginal distribution 
of the unordered singular values of $\boldsymbol{H}^{k}$ \cite{Tulino04}. 
\end{corollary}

The above result for the STS scheme is consistent with that 
reported in \cite{Hanly01,Mantravadi03}. 
On the other hand, the result for the TS scheme includes and extends 
the result in \cite{Mantravadi03}, the latter of which has been proved 
only when the number of receive antennas is one or when the number 
of transmit antennas is infinite.

\section{Numerical Results} \label{section_numerical}
\begin{center}
 \begin{figure}[tbp]
  \includegraphics[width=\hsize]{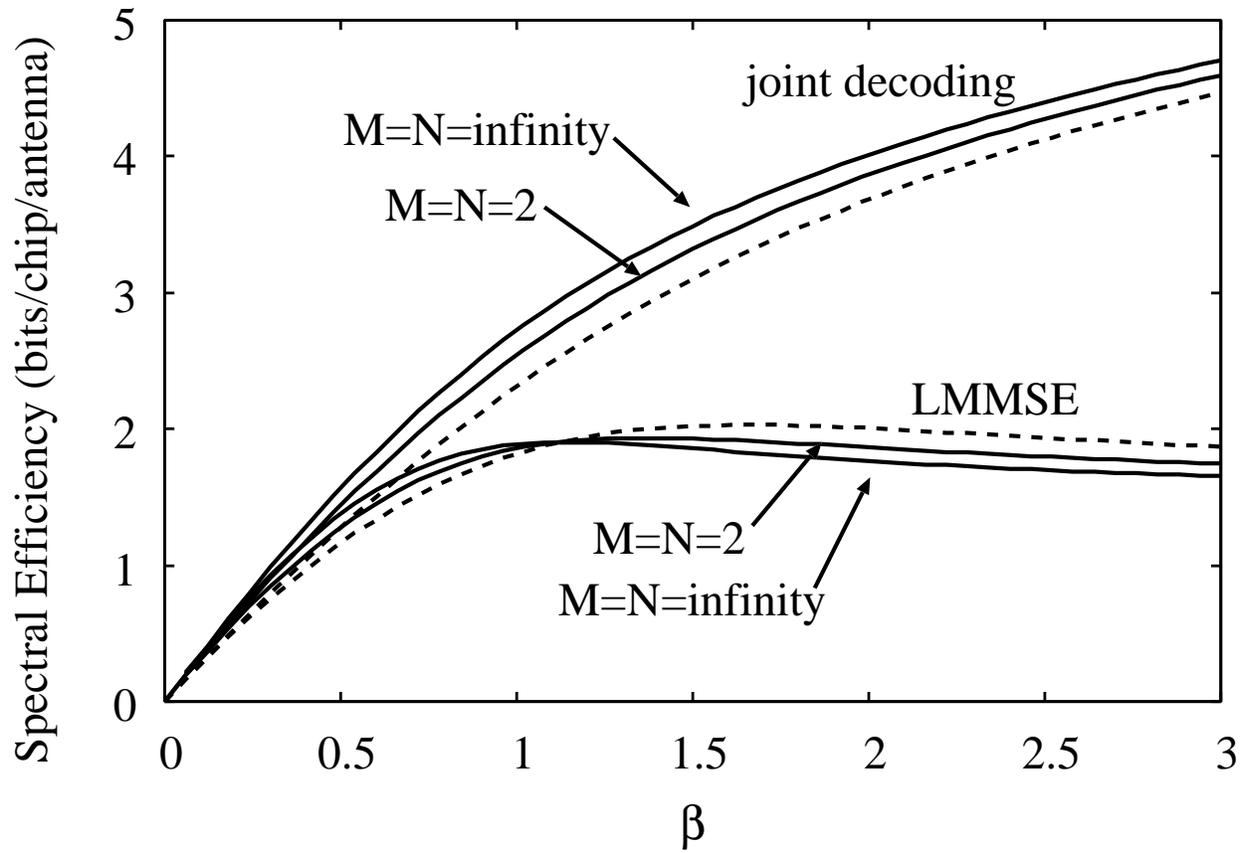}
  \caption{
  Spectral efficiency per transmit antenna of the MIMO DS-CDMA channel 
  versus $\beta$ in the case of the Gaussian data modulation. 
  The solid lines are the spectral efficiencies per transmit antenna 
  in the case of the STS scheme. The dashed lines 
  represent the spectral efficiencies per transmit antenna in the case 
  of the TS scheme. The dashed lines overlap for 
  $M=N=2,\ldots,\infty$. $P/N_{0}=10$ dB. 
  }
  \label{gauss}
 \end{figure}
\end{center}

We compare numerically the spectral efficiency of the MIMO DS-CDMA channel 
in the case of several transmitter and receiver structures. In this section,  
we assume that the number of transmit antennas is uniform, i.e., 
$M_{k}=M$, that $\{x_{m}^{k}\}$ are i.i.d. random variables with   
variance $P$, and $\{h_{nm}^{k}\}\sim$ i.i.d. $\mathcal{CN}(0,1/N)$. 
From these assumptions, the spectral 
efficiencies~(\ref{spectral_efficiency_rs_SIMO}), 
(\ref{spectral_efficiency_rs_MIMO}), 
(\ref{spectral_efficiency_separate_SIMO}), 
and~(\ref{spectral_efficiency_separate_MIMO}) converge almost surely 
to their expectations with respect to $\boldsymbol{\mathcal{H}}$. 
Figure~\ref{gauss} shows the spectral efficiency of the MIMO DS-CDMA channel 
with the STS scheme and that with the TS scheme,  
in the case of the Gaussian data modulation.  
The maximum spectral efficiency of the MIMO DS-CDMA channel with the 
STS scheme is slightly larger than that with the TS scheme even if 
optimal space-time encoding is employed. 
This result implies that the TS scheme is suboptimal.  
On the other hand, the spectral efficiency of the 
MIMO DS-CDMA channel with the MMSE detector front end is slightly larger 
in the case of the TS scheme than in the case of the 
STS scheme when the system load $\beta$ is high. 

\begin{center}
 \begin{figure}[tbp]
  \includegraphics[width=\hsize]{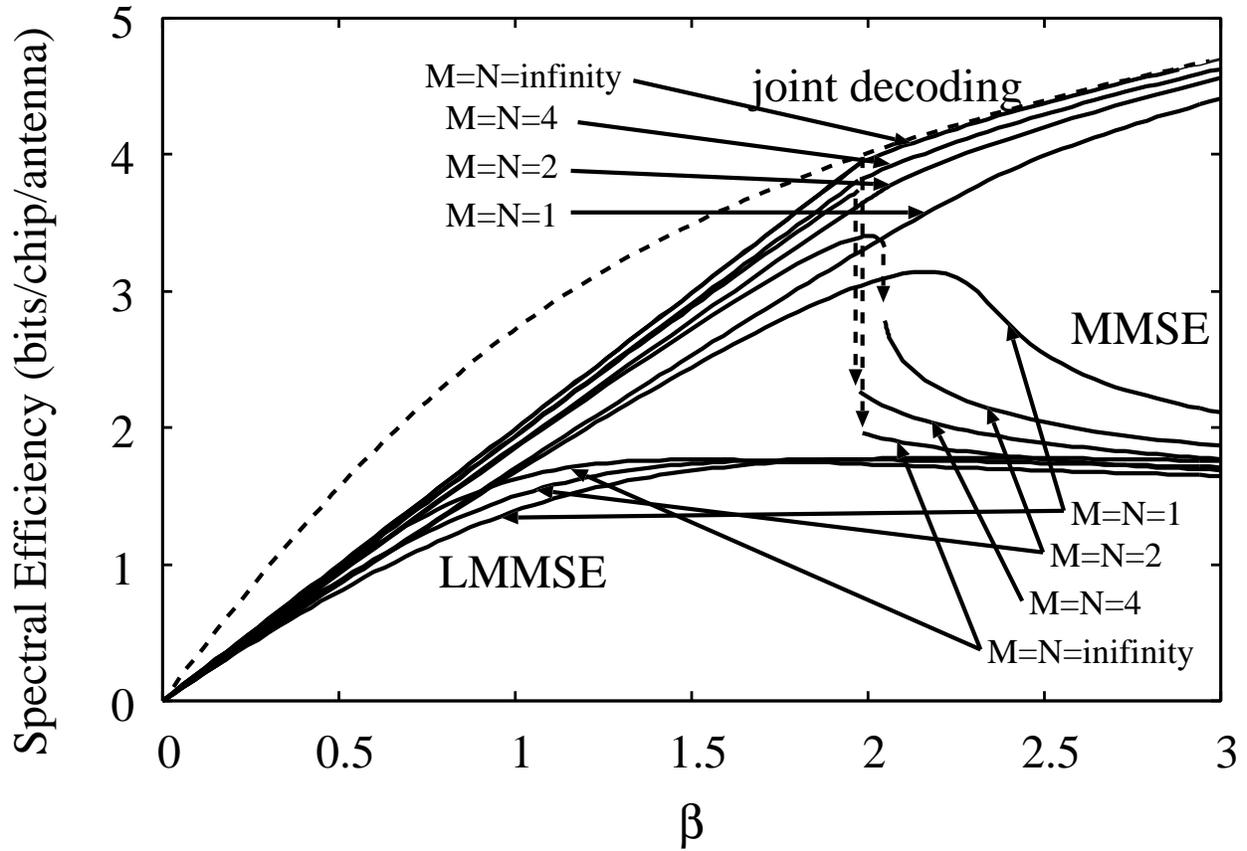}
  \caption{
  Spectral efficiency per transmit antenna of the MIMO DS-CDMA channel 
  with the QPSK data modulation scheme versus $\beta$ in the case of 
  the STS scheme. The dashed line 
  is the maximum spectral efficiency per transmit antenna of the MIMO DS-CDMA 
  channel with the Gaussian data modulation scheme and the STS scheme 
  for $M=N=\infty$. $P/N_{0}=10$ dB. 
  }
  \label{qpsk}
 \end{figure}
\end{center}

We next consider the spectral efficiency of the MIMO DS-CDMA channel 
in the case in which the data modulation scheme is specified. 
Figure~\ref{qpsk} shows the spectral efficiency of the MIMO DS-CDMA channel 
with the QPSK data modulation scheme in the case of the STS  
scheme. The spectral efficiency of the MIMO DS-CDMA channel 
with the MMSE detector front end displays {\it waterfall} behavior in 
the case of multiple antennas. The spectral efficiency is much larger 
than the spectral efficiency of the MIMO DS-CDMA channel with the LMMSE 
detector when the system load $\beta$ 
is smaller than the transition point of the {\it waterfall} behavior. 
Furthermore, the spectral efficiency is comparable to the maximum 
spectral efficiency of the MIMO DS-CDMA channel, which is very close to the   
sum capacity when the system load is just below the transition point.  
However, it decreases rapidly as the system load $\beta$ becomes larger 
than the transition point. 

\section{Discussion} \label{section_extension}
The results in Section~\ref{section_numerical} imply that the separation 
loss can be reduced dramatically by employing a non-Gaussian data modulation 
scheme, even though it has less information per received power than the 
Gaussian data modulation, and taking the data modulation scheme into 
consideration in multiuser detection. 
Although it is difficult to perform the multiuser detection exactly, 
approximate methods of multiuser detection, e.g., an algorithm on the 
basis of message passing, may provide higher performance than 
linear multiuser detectors. 

One can obtain, from Propositions~\ref{Claim_decoupling} 
and~\ref{claim_separate}, hints about an appropriate structure 
of the transmitter and a method of constructing multiuser detection 
algorithms in many-user systems.  
In the case of the STS scheme, the spectral efficiency of the MIMO DS-CDMA 
channel with the MMSE detector front end is achieved by per-antenna encoding 
because (\ref{spectral_efficiency_separate_SIMO}) is given by the sum of 
the spectral efficiencies of the SIMO Gaussian channel~(\ref{SIMO_channel}) 
with the MMSE detector front end.  
This means that space-time encoding is not required, and furthermore, that 
the ergodic spectral efficiency 
$\mathrm{E}[\mathcal{C}_{\mathrm{sep}}^{\mathrm{STS}}]$ does not decrease 
even if the transmit spatial correlation at each user exists.  
Another suggestion is that one should take a space-time encoded symbol 
vector $\boldsymbol{x}^{k}$ as a single node in a message passing algorithm 
for the TS scheme, in order to acquire comparable spectral efficiency to 
that in the case of the STS scheme, although one can take the element 
$x_{m}^{k}$ of the symbol vector as a single node for the STS scheme.

Although we have so far considered the uplink of the MIMO DS-CDMA 
channel, our results can straightforwardly be extended to the cases of 
an uplink of a frequency-selective MIMO channel with the MC CDMA scheme 
(MIMO MC-CDMA) \cite{Juntti05} and downlinks of these channels. 

Our model corresponds to the uplink of the MIMO MC-CDMA channel 
when $\boldsymbol{H}^{k}$ is replaced by 
$\boldsymbol{H}_{l}^{k}$, which represents the channel matrix of 
the $l$th subcarrier from the $k$th user to the receiver. 
Furthermore, (\ref{mutual_information}) coincides with the 
maximum spectral efficiency per chip and per sub-carrier of the 
MIMO MC-CDMA channel without encoding across the subcarriers if the loss of 
spectral efficiency due to inserting a cyclic prefix is neglected. 
If $\{\boldsymbol{H}_{l}^{k};l=1.\ldots,L\}$ are i.i.d., 
Propositions~\ref{claim_spectral_efficiency}--\ref{claim_separate},  
in which $\boldsymbol{H}^{k}$ is replaced by $\boldsymbol{H}_{1}^{k}$, 
hold for the case in which the ergodic spectral efficiency is considered. 
On the other hand, our model corresponds to the downlinks of the   
MIMO DS-CDMA channel and the MIMO MC-CDMA channel 
if $\{\boldsymbol{H}^{k};k=1,\ldots,K\}$ and 
$\{\boldsymbol{H}_{l}^{k};k=1,\ldots,K\}$, respectively, represent  
the same fading \cite{Guo06}. In addition, in this case, 
Propositions~\ref{claim_spectral_efficiency}--\ref{claim_separate} hold 
without modification. 
It is easy to extend these results to the case with encoding across 
the subcarriers.

\section{Conclusion} \label{section_conclusion}
We have analyzed the MIMO DS-CDMA channel with the GPME detector front end. 
In the many-user limit, the MIMO DS-CDMA channel with the MMSE detector 
front end and the STS scheme is decoupled into a bank of the single-user 
SIMO Gaussian channels with the MMSE detector front ends. 
On the other hand, the MIMO DS-CDMA channel with the MMSE detector front end 
and the TS scheme is decoupled into a bank of the single-user MIMO 
Gaussian channels with the MMSE detector front end. 
If suitable space-time encoding is employed, the spectral efficiency 
of the MIMO DS-CDMA channel with the TS scheme is comparable 
with the spectral efficiency of the MIMO DS-CDMA channel with the STS 
scheme, where space-time encoding is not required. 
Our results suggest huge potentialities of nonlinear multiuser 
detection, which can exceed linear multiuser detection, and a guideline 
for methods of constructing message passing algorithms for multiuser 
detection in many-user systems.

\appendices
\section{Derivation of Proposition~\ref{claim_spectral_efficiency}} 
\label{appendix_free_energy} 
We define the free energy $\mathcal F$ as
\begin{equation}
{\mathcal F} = 
-\lim_{K,L\rightarrow \infty}\frac{1}{K}
\mathrm{E}\left\{
 \left.
  \ln \mathrm{E}_{\vec{\tilde{\boldsymbol{x}}}}
  \left[
   p(\vec{\tilde{\boldsymbol{y}}}=\vec{\boldsymbol{y}}|
   \vec{\tilde{\boldsymbol{x}}}, \boldsymbol{\mathcal S}, 
   \boldsymbol{\mathcal H};\tilde{N}_{0})
  \right]
 \right| \boldsymbol{\mathcal H}
\right\}, \label{free_energy} 
\end{equation}
where the limit in (\ref{free_energy}) stands for the many-user limit. 
From Assumption~\ref{assumption_self-averaging}, in the many-user limit, 
the maximum spectral efficiency~(\ref{mutual_information}) is given by 
\begin{equation} \label{spectral_efficiency}
\lim_{K,L\rightarrow\infty}{\mathcal C}_{\mathrm{joint}} = 
\frac{\beta}{\ln 2}\left.
 {\mathcal F}
\right|_{p(\vec{\tilde{\boldsymbol{x}}}) =
p(\vec{\boldsymbol{x}}), \tilde{N}_{0} = N_{0}}
- N\log(\pi N_{0}\mathrm{e}).  
\end{equation}  

We evaluate the free energy~(\ref{free_energy}) using the replica trick. 
Applying to (\ref{free_energy}) the identity
\begin{equation}
\mathrm{E}[\ln Z] =
\lim_{u\rightarrow 0}\frac{\partial}{\partial u}\ln\mathrm{E}[Z^{u}] 
\quad {\rm for}\;Z>0\;a.s.,  
\end{equation}
we have  
\begin{subequations}
\begin{align}
{\mathcal F} &= 
-\lim_{u\rightarrow 0}\frac{\partial}{\partial u}
\lim_{K,L\rightarrow \infty}\frac{1}{K}
\ln\Xi^{(u)}, \label{free_energy_replica}  \\
\Xi^{(u)} &= \mathrm{E}\left[
 \left.
  \left\{
   \mathrm{E}_{\vec{\tilde{\boldsymbol{x}}}}
   \left[
    p(\vec{\tilde{\boldsymbol{y}}}=\vec{\boldsymbol{y}}|
    \vec{\tilde{\boldsymbol{x}}}, \boldsymbol{\mathcal S}, 
    \boldsymbol{\mathcal H};\tilde{N}_{0})
   \right]
  \right\}^{u}
 \right| \boldsymbol{\mathcal H} 
\right], \label{Xi} 
\end{align}
\end{subequations}
where we have exchanged the order of the many-user limit and the operations 
with respect to $u$. 
We evaluate (\ref{Xi}) on $\{u=0,1,\ldots\}$ in the many-user limit, 
whose domain is extended to the real number field in the 
neighborhood of $u=0$. The validity of these procedures is pending. 

We write the original symbol vector for the $k$th user and 
the i.i.d. replicated symbol vectors for the $k$th user following 
$p(\tilde{\boldsymbol{x}}^{k})$ as $\boldsymbol{x}^{k,0}$ and 
$\boldsymbol{x}^{k,\alpha}=(x_{1}^{k,\alpha}, \ldots, 
x_{M_{k}}^{k,\alpha})^{T}$ for $\alpha=1,\ldots,u$, respectively. 
Due to the independence of the spreading sequences,  
(\ref{Xi}) yields 
\begin{subequations}
\begin{align}
\Xi^{(u)} &= 
\mathrm{E}\left\{
 \left.
  \exp\left[
   LG_{K}^{(u)}(\vec{\boldsymbol{\mathcal X}})
  \right] 
 \right| \boldsymbol{\mathcal H}
\right\}, \label{tmp2} \\
G_{K}^{(u)}(\vec{\boldsymbol{\mathcal X}}) &=  
\ln \mathrm{E}\left[
 \left.  
  \int\prod_{\alpha =0}^{u}\exp\left(
   -\frac{1}{N_{\alpha}}\left\|
    \boldsymbol{y} - \sqrt{\beta}\boldsymbol{v}^{\alpha} 
   \right\|^{2}
  \right) 
  \boldsymbol{dy} \right| 
 \vec{\boldsymbol{\mathcal X}}, \boldsymbol{\mathcal H}  
\right] 
-N\sum_{\alpha =0}^{u}\ln(\pi N_{\alpha}), \label{G_K}
\end{align}
\end{subequations}
where $\vec{\boldsymbol{\mathcal X}}=
\{\boldsymbol{x}^{k,\alpha}; k=1,\ldots,K, \alpha =0, \ldots ,u\}$, 
$N_{\alpha}=\tilde{N}_{0}$ for $\alpha=1, \ldots, u$ and    
$N$-dimensional vector $\boldsymbol{v}^{\alpha}$ is defined as  
\begin{equation}
\boldsymbol{v}^{\alpha} = 
\frac{1}{\sqrt{\beta}}\sum_{k=1}^{K}\boldsymbol{H}^{k}
\boldsymbol{S}^{k}\boldsymbol{x}^{k,\alpha}
\end{equation}
with i.i.d. $M_{k}\times M_{k}$ random matrices 
$\{\boldsymbol{S}^{k};k=1, \ldots, K\}$ following $p(\boldsymbol{S}_{l}^{k})$.
Since $K$ and $L$ are sufficiently large with their ratio 
$\beta=K/L$ fixed, due to the central limit theorem,  
$\boldsymbol{v} = ({\boldsymbol{v}^{0}}^{T}, \ldots , 
{\boldsymbol{v}^{u}}^{T})^{T}$ conditioned on 
$\vec{\boldsymbol{\mathcal X}}$ and $\boldsymbol{\mathcal H}$ 
follows approximately the zero-mean circularly symmetric complex Gaussian 
distribution with the covariance matrix 
\begin{equation}
\boldsymbol{\mathcal Q} = 
\frac{1}{K}\sum_{k=1}^{K}\left[
 \delta\boldsymbol{w}^{k}(\boldsymbol{w}^{k})^{H}
 + (1-\delta)\boldsymbol{W}^{k}(\boldsymbol{W}^{k})^{H} 
\right], \label{covariance_matrix} 
\end{equation}
where $\boldsymbol{w}^{k}= [(\boldsymbol{H}^{k}\boldsymbol{x}^{k,0})^{T} 
\cdots (\boldsymbol{H}^{k}\boldsymbol{x}^{k,u})^{T}]^{T}$ and 
$\boldsymbol{W}^{k}= [(\boldsymbol{W}^{k,0})^{T} \cdots 
(\boldsymbol{W}^{k,u})^{T}]^{T}$ with 
$\boldsymbol{W}^{k,\alpha} = (\boldsymbol{h}_{1}^{k}x_{1}^{k,\alpha} \cdots  
\boldsymbol{h}_{M_{k}}^{k}x_{M_{k}}^{k,\alpha})$.  
Evaluating the Gaussian integration in (\ref{G_K}), we obtain     
\begin{subequations}
\begin{align}
\frac{1}{K}\ln\Xi^{(u)} &= 
\frac{1}{K}\ln\mathrm{E}\left\{
 \left. 
  \exp\left[ 
   K\beta^{-1}G^{(u)}(\boldsymbol{\mathcal Q})
  \right]
 \right| \boldsymbol{\mathcal H}
\right\} + \mathcal{O}(K^{-1}), \label{tmp4} \\
G^{(u)}(\boldsymbol{\mathcal Q}) &= 
-\ln\det(\boldsymbol{I}+\boldsymbol{\Sigma}\boldsymbol{\mathcal Q}) 
-Nu\ln (\pi\tilde{N}_{0}) 
-N\ln\left(
 1+\frac{N_{0}}{\tilde{N}_{0}}u
\right), \label{G}
\end{align}
\end{subequations}
where $\boldsymbol{\Sigma}$ is given by 
\begin{equation}
\boldsymbol{\Sigma} = 
\frac{\beta}{\tilde{N}_{0}+uN_{0}}
\begin{bmatrix}
 u & -\boldsymbol{e}_{u}^{T} \\
 -\boldsymbol{e}_{u} & \left(1+\frac{uN_{0}}{\tilde{N}_{0}}\right)
 \boldsymbol{I}_{u} - \frac{N_{0}}{\tilde{N}_{0}}
 \boldsymbol{e}_{u}\boldsymbol{e}_{u}^{T}  
\end{bmatrix}
\otimes \boldsymbol{I}_{N}. 
\end{equation}

The expectation with respect to $\vec{\boldsymbol{\mathcal X}}$ in 
(\ref{tmp4}) is evaluated as the expectation with respect to 
$\boldsymbol{\mathcal Q}$ 
because (\ref{G}) depends on $\vec{\boldsymbol{\mathcal X}}$ only through 
$\boldsymbol{\mathcal Q}$.
Since $\boldsymbol{\mathcal Q}$ is the empirical mean of independent 
random variables conditioned on $\boldsymbol{\mathcal H}$,  
the probability density function of $\boldsymbol{\mathcal Q}$ 
conditioned on $\boldsymbol{\mathcal H}$   
satisfies the large deviation principle
\begin{equation}
\frac{1}{K}\ln p_{K}^{(u)}(\boldsymbol{\mathcal Q}|\boldsymbol{\mathcal H}) = 
-I^{(u)}(\boldsymbol{\mathcal Q}) + o(K), 
\end{equation}  
where the rate function $I^{(u)}(\boldsymbol{\mathcal Q})$ is given by
\begin{equation}
I^{(u)}(\boldsymbol{\mathcal Q}) = 
\sup_{\tilde{\boldsymbol{\mathcal Q}}}\left[
 \mathrm{Tr}(\tilde{\boldsymbol{\mathcal Q}}\boldsymbol{\mathcal Q}) 
 - \lim_{K\rightarrow\infty}\frac{1}{K}\sum_{k=1}^{K}
 \ln{\mathcal M}_{k}^{(u)}(\tilde{\boldsymbol{\mathcal Q}}) 
\right] \label{rate_function}
\end{equation} 
with the moment generating function of the symbols of the $k$th user 
${\mathcal M}_{k}^{(u)}(\tilde{\boldsymbol{\mathcal Q}})$ defined as  
\begin{equation}
{\mathcal M}_{k}^{(u)}(\tilde{\boldsymbol{\mathcal Q}}) =
\mathrm{E}\left\{
 \left. 
  \mathrm{e}^{
   \Lambda^{k}(\tilde{\boldsymbol{\mathcal Q}})
  }
 \right| \boldsymbol{H}^{k}
\right\}, 
\quad
\Lambda^{k} = 
\delta\mathrm{Tr}\left(
 \tilde{\boldsymbol{\mathcal Q}}\boldsymbol{w}^{k}(\boldsymbol{w}^{k})^{H} 
\right)
+ (1-\delta)\mathrm{Tr}\left(
 \tilde{\boldsymbol{\mathcal Q}}\boldsymbol{W}^{k}(\boldsymbol{W}^{k})^{H} 
\right). \label{moment_generating}
\end{equation}
Evaluating the expectation in (\ref{tmp4}) with the saddle point method, 
we have    
\begin{equation}
\lim_{K,L\rightarrow\infty}\frac{1}{K}\ln\Xi^{(u)} = 
\sup_{\boldsymbol{\mathcal Q}}\left[
 \beta^{-1}G^{(u)}(\boldsymbol{\mathcal Q}) 
 - I^{(u)}(\boldsymbol{\mathcal Q})
\right]. \label{tmp5}
\end{equation} 
Setting the derivatives of (\ref{rate_function}) and 
(\ref{tmp5}) with respect to $\tilde{\boldsymbol{\mathcal Q}}$ and 
$\boldsymbol{\boldsymbol{Q}}$, respectively, to zero, 
we obtain the following equations, which give extrema of 
(\ref{rate_function}) and (\ref{tmp5}):  
\begin{subequations}
\begin{align} 
\boldsymbol{\mathcal Q}^{\mathrm{s}} &=  
\lim_{K\rightarrow\infty}\frac{1}{K}\sum_{k=1}^{K} 
\frac{1}{{\mathcal M}_{k}^{(u)}
(\tilde{\boldsymbol{\mathcal Q}}^{\mathrm{s}})}
\mathrm{E}\left\{
 \left.  
  \left[
   \delta\boldsymbol{w}^{k}(\boldsymbol{w}^{k})^{H}
   + (1-\delta)\boldsymbol{W}^{k}(\boldsymbol{W}^{k})^{H}
  \right]
  \mathrm{e}^{
   \Lambda^{k}(\tilde{\boldsymbol{\mathcal Q}}^{\mathrm{s}})
  }
 \right| \boldsymbol{H}^{k}
\right\}, \label{equation1} \\ 
\tilde{\boldsymbol{\mathcal Q}}^{\mathrm{s}} &=
-\beta^{-1}(\boldsymbol{I}+\boldsymbol{\Sigma}
\boldsymbol{\mathcal Q}^{\mathrm{s}})^{-1}\boldsymbol{\Sigma}. 
\label{equation2}
\end{align}
\end{subequations}
Substituting (\ref{equation2}) into the derivative of (\ref{tmp5}) 
with respect to $u$, we can evaluate (\ref{free_energy_replica}) as 
\begin{equation}
{\mathcal F} = 
-\lim_{u\rightarrow 0}\left[
 \beta^{-1}\frac{\partial G^{(u)}}{\partial u}
 (\boldsymbol{\mathcal Q}^{\mathrm{s}}) 
 - \frac{\partial I^{(u)}}{\partial u}
 (\boldsymbol{\mathcal Q}^{\mathrm{s}})  
\right]. \label{tmp6}
\end{equation} 

We have to solve (\ref{equation1}) and (\ref{equation2}) analytically 
in order to evaluate (\ref{tmp6}).  
In what follows, we assume that the replica symmetry 
holds\footnote{It should be noted that there exist cases in which the 
replica symmetry is not valid. 
See \cite{Yoshida07} for a discussion of the validity 
in the context of DS-CDMA.}, i.e., 
$\boldsymbol{\mathcal Q}^{\mathrm{s}}$ and 
$\tilde{\boldsymbol{\mathcal Q}}^{\mathrm{s}}$ are invariant 
under the exchange of non-zero replica indexes. 
Then, $\boldsymbol{\mathcal Q}^{\mathrm{s}}$ and 
$\tilde{\boldsymbol{\mathcal Q}}^{\mathrm{s}}$ in the limit $u\rightarrow 0$  
can be written as 
\begin{subequations}
\begin{align}
\boldsymbol{\mathcal Q}^{\mathrm{s}} &= 
\begin{pmatrix}
 \boldsymbol{Q}^{0} & \boldsymbol{e}_{u}^{T}\otimes \boldsymbol{M} \\
 \boldsymbol{e}_{u}\otimes\boldsymbol{M}^{H} 
 & \boldsymbol{I}_{u}\otimes (\boldsymbol{Q}^{1}-\boldsymbol{Q}) 
 + \boldsymbol{e}_{u}\boldsymbol{e}_{u}^{T}\otimes\boldsymbol{Q}
\end{pmatrix}, \\
\tilde{\boldsymbol{\mathcal Q}}^{\mathrm{s}} &= 
\begin{pmatrix}
 \tilde{\boldsymbol{Q}}^{0} & 
 \boldsymbol{e}_{u}^{T}\otimes\tilde{\boldsymbol{M}} \\
 \boldsymbol{e}_{u}\otimes\tilde{\boldsymbol{M}}^{H} 
 & \boldsymbol{I}_{u}\otimes (\tilde{\boldsymbol{Q}}^{1}
 -\tilde{\boldsymbol{Q}}) 
 + \boldsymbol{e}_{u}\boldsymbol{e}_{u}^{T}\otimes
 \tilde{\boldsymbol{Q}}
\end{pmatrix}
\end{align}
\end{subequations}
with $N\times N$ matrices $\boldsymbol{M}$, $\tilde{\boldsymbol{M}}$ 
and $N\times N$ Hermitian matrices 
$\boldsymbol{Q}^{0}$, $\tilde{\boldsymbol{Q}}^{0}$, 
$\boldsymbol{Q}^{1}$, $\tilde{\boldsymbol{Q}}^{1}$, 
$\boldsymbol{Q}$, and $\tilde{\boldsymbol{Q}}$.  
Comparing both sides of (\ref{equation2}), we obtain 
in the limit $u\rightarrow 0$ 
\begin{equation}
\tilde{\boldsymbol{Q}}^{0} = 
\boldsymbol{0}, 
\quad
\tilde{\boldsymbol{M}} = 
\tilde{\boldsymbol{R}}^{-1},
\quad 
\tilde{\boldsymbol{Q}}^{1} = 
\tilde{\boldsymbol{Q}} 
- \tilde{\boldsymbol{M}}, 
\quad
\tilde{\boldsymbol{Q}} = 
\tilde{\boldsymbol{R}}^{-1}\boldsymbol{R}\tilde{\boldsymbol{R}}^{-1}, 
\label{solution2}
\end{equation}
where $\boldsymbol{R}$ and $\tilde{\boldsymbol{R}}$ are respectively given by 
\begin{subequations}
\begin{align}
\boldsymbol{R} &= 
N_{0}\boldsymbol{I}_{N} 
+ \beta(\boldsymbol{Q}^{0}-\boldsymbol{M}-\boldsymbol{M}^{H}
+\boldsymbol{Q}), \label{R_0_tmp} \\ 
\tilde{\boldsymbol{R}} &= 
\tilde{N}_{0}\boldsymbol{I}_{N} 
+ \beta(\boldsymbol{Q}^{1}-\boldsymbol{Q}). \label{R_tmp}
\end{align}
\end{subequations}

Next, We solve (\ref{equation1}). We describe only the  
evaluation of the moment generating function~(\ref{moment_generating}) 
because each term of (\ref{equation1}) is evaluated in the same manner. 
It is straightforward to prove that $\boldsymbol{R}$ is positive definite.  
Then, one can take a square root of $\boldsymbol{R}$ such that 
$\boldsymbol{R}=\sqrt{\boldsymbol{R}}(\sqrt{\boldsymbol{R}})^{H}$. 
Substituting (\ref{solution2}) into (\ref{moment_generating}), we obtain  
\begin{eqnarray}
{\mathcal M}_{k}^{(u)}(\tilde{\boldsymbol{\mathcal Q}}^{\mathrm{s}}) &=& 
\mathrm{E}\left\{
 \exp\left[
  \delta\left(
   \left\| 
    \boldsymbol{b}^{k} 
   \right\|^{2} 
   - \sum_{\alpha =0}^{u}(\boldsymbol{H}^{k}\boldsymbol{x}^{k,\alpha})^{H}
   \boldsymbol{R}_{\alpha}^{-1}\boldsymbol{H}^{k}\boldsymbol{x}^{k,\alpha} 
  \right)
 \right.
\right. \nonumber \\
&& 
\left.
 \left.
  \left. 
   + (1-\delta)\sum_{m=1}^{M_{k}}\left(
    \left\|
     \boldsymbol{b}_{m}^{k}
    \right\|^{2}  
    - \sum_{\alpha =0}^{u}(\boldsymbol{h}_{m}^{k}
    x_{m}^{k,\alpha})^{H}
    \boldsymbol{R}_{\alpha}^{-1}\boldsymbol{h}_{m}^{k}x_{m}^{k,\alpha}
   \right)
  \right]
 \right| \boldsymbol{H}^{k}
\right\}, \label{moment_generating_tmp}
\end{eqnarray}
where $\boldsymbol{R}_{0}=\boldsymbol{R}$, 
$\boldsymbol{R}_{\alpha}=\tilde{\boldsymbol{R}}$ for $\alpha=1, \ldots, u$
and $\boldsymbol{b}^{k}$ and $\boldsymbol{b}_{m}^{k}$ are given by 
\begin{subequations}
\begin{align} 
\boldsymbol{b}^{k} &= 
(\sqrt{\boldsymbol{R}})^{-1}\boldsymbol{H}^{k}\boldsymbol{x}^{k,0} 
+ (\sqrt{\boldsymbol{R}})^{H}\tilde{\boldsymbol{R}}^{-1}\sum_{\alpha =1}^{u}
\boldsymbol{H}^{k}\boldsymbol{x}^{k,\alpha}, \\
\boldsymbol{b}_{m}^{k} &= 
(\sqrt{\boldsymbol{R}})^{-1}\boldsymbol{h}_{m}^{k}x_{m}^{k,0} 
+ (\sqrt{\boldsymbol{R}})^{H}\tilde{\boldsymbol{R}}^{-1}\sum_{\alpha =1}^{u}
\boldsymbol{h}_{m}^{k}x_{m}^{k,\alpha}.  
\end{align}
\end{subequations}
Substituting two transforms that are similar to the Hubbard-Stratonovich 
transform, as follows:
\begin{subequations}
\begin{align}
\mathrm{e}^{\delta|\boldsymbol{b}^{k}\|^{2}} &= 
\int \frac{1}{\pi^{N}\det (\delta^{-1}\boldsymbol{R})}
\mathrm{e}^{
 \delta\left[
  -(\boldsymbol{y}^{k})^{H}\boldsymbol{R}^{-1}\boldsymbol{y}^{k} 
  + 2\mathrm{Re}((\boldsymbol{b}^{k})^{H}(\sqrt{\boldsymbol{R}})^{-1}
  \boldsymbol{y}^{k})
 \right]
}\boldsymbol{dy}^{k}, \\
\mathrm{e}^{(1-\delta)\|\boldsymbol{b}_{m}^{k}\|^{2}} &= 
\int \frac{1}{\pi^{N}\det[(1-\delta)^{-1}\boldsymbol{R}]}
\mathrm{e}^{
 (1-\delta)\left[
  -(\boldsymbol{y}_{m}^{k})^{H}\boldsymbol{R}^{-1}
  \boldsymbol{y}_{m}^{k} 
  + 2\mathrm{Re}((\boldsymbol{b}_{m}^{k})^{H}
  (\sqrt{\boldsymbol{R}})^{-1}\boldsymbol{y}_{m}^{k})
 \right]
}\boldsymbol{dy}_{m}^{k}, 
\end{align}
\end{subequations}
into (\ref{moment_generating_tmp}), we have  
\begin{equation} \label{moment_generating_rs}
{\mathcal M}_{k}^{(u)}
(\tilde{\boldsymbol{\mathcal Q}}^{\mathrm{s}}) 
=
\int \mathrm{E}_{\boldsymbol{x}^{k}}[
 q(
  \boldsymbol{y}^{k}, \{\boldsymbol{y}_{m}^{k}\} |
   \boldsymbol{x}^{k}, \boldsymbol{H}^{k}; \boldsymbol{R}
 )
]
\left\{
 \frac{
  \mathrm{E}_{\tilde{\boldsymbol{x}}^{k}}[
   q(
    \boldsymbol{y}^{k}, \{\boldsymbol{y}_{m}^{k}\} |
     \tilde{\boldsymbol{x}}^{k}, \boldsymbol{H}^{k};\tilde{\boldsymbol{R}}
   )
  ]
 }
 {
  q(
   \boldsymbol{y}^{k}, \{\boldsymbol{y}_{m}^{k}\} |
    \boldsymbol{0}, \boldsymbol{H}^{k};\tilde{\boldsymbol{R}}
  )
 }
\right\}^{u}\boldsymbol{dy}^{k}\prod_{m=1}^{M_{k}}\boldsymbol{dy}_{m}^{k}, 
\end{equation}
where the function $q(\boldsymbol{y}^{k},\{\boldsymbol{y}_{m}^{k}\}|
\boldsymbol{x}^{k}, \boldsymbol{H}^{k}; \boldsymbol{R})$ is given by 
\begin{equation}
q(
 \boldsymbol{y}^{k}, \{\boldsymbol{y}_{m}^{k}\} |
  \boldsymbol{x}^{k}, \boldsymbol{H}^{k}; \boldsymbol{R}
) = 
q(
 \boldsymbol{y}^{k} |
  \boldsymbol{x}^{k}, \boldsymbol{H}^{k}; \delta^{-1}\boldsymbol{R}
)
\prod_{m=1}^{M_{k}}p(
  \boldsymbol{y}_{m}^{k} |
 x_{m}^{k}, \boldsymbol{h}_{m}^{k}; 
 (1-\delta)^{-1}\boldsymbol{R}
), 
\end{equation}
\begin{subequations}
\begin{align}
q(
 \boldsymbol{y}^{k} |
  \boldsymbol{x}^{k}, \boldsymbol{H}^{k}; \boldsymbol{R}
) &= 
\frac{1}{\pi^{N}\det\boldsymbol{R}}\exp\left[ 
 - (\boldsymbol{y}^{k}-\boldsymbol{H}^{k}\boldsymbol{x}^{k})^{H}
 \boldsymbol{R}^{-1}
 (\boldsymbol{y}^{k}-\boldsymbol{H}^{k}\boldsymbol{x}^{k})
\right], \\
q(
  \boldsymbol{y}_{m}^{k} |
 x_{m}^{k}, \boldsymbol{h}_{m}^{k}; \boldsymbol{R}
) &= 
\frac{1}{\pi^{N}\det\boldsymbol{R}}
\exp\left[ 
 - (\boldsymbol{y}_{m}^{k}-\boldsymbol{h}_{m}^{k}x_{m}^{k})^{H}
 \boldsymbol{R}^{-1}
 (\boldsymbol{y}_{m}^{k}-\boldsymbol{h}_{m}^{k}x_{m}^{k})
\right].
\end{align}
\end{subequations}

Evaluating (\ref{equation1}) in the same manner and substituting the results 
into (\ref{R_0_tmp}) and (\ref{R_tmp}), we obtain the following 
fixed-point equations: 
\begin{subequations}
\begin{align}
\boldsymbol{R} &= 
N_{0}\boldsymbol{I}_{N} 
+ \beta\lim_{K\rightarrow\infty}\frac{1}{K}\sum_{k=1}^{K}\left[
 \delta\boldsymbol{H}^{k}\boldsymbol{\mathcal E}_{\mathrm{MIMO}}^{k,\delta} 
 (\boldsymbol{H}^{k})^{H}
 + (1-\delta)\sum_{m=1}^{M_{k}}
 {\mathcal E}_{\mathrm{SIMO}}^{k,m,\delta}
 \boldsymbol{h}_{m}^{k}(\boldsymbol{h}_{m}^{k})^{H}
\right], \label{R_0_apen} \\ 
\tilde{\boldsymbol{R}} &=
\tilde{N}_{0}\boldsymbol{I}_{N} 
+ \beta\lim_{K\rightarrow\infty}\frac{1}{K}\sum_{k=1}^{K}\left[
 \delta\boldsymbol{H}^{k}\boldsymbol{\mathcal V}_{\mathrm{MIMO}}^{k,\delta}
 (\boldsymbol{H}^{k})^{H} 
 + (1-\delta)\sum_{m=1}^{M_{k}}
 {\mathcal V}_{\mathrm{SIMO}}^{k,m,\delta}
 \boldsymbol{h}_{m}^{k}(\boldsymbol{h}_{m}^{k})^{H}
\right] \label{R_apen}
\end{align}
\end{subequations}
with $\boldsymbol{\mathcal E}_{\mathrm{MIMO}}^{k,\delta}$, 
${\mathcal E}_{\mathrm{SIMO}}^{k,\delta}$, 
$\boldsymbol{\mathcal V}_{\mathrm{MIMO}}^{k,m,\delta}$, and 
${\mathcal V}_{\mathrm{SIMO}}^{k,m,\delta}$ 
defined respectively as  
\begin{subequations}
\begin{align}
\boldsymbol{\mathcal E}_{\mathrm{MIMO}}^{k,\delta} &=
\mathrm{E}\left[
 \left.
  \left(
   \boldsymbol{x}^{k} 
   - \langle \tilde{\boldsymbol{x}}^{k} \rangle_{\mathrm{s}} 
  \right)
  \left(
   \boldsymbol{x}^{k} 
   - \langle \tilde{\boldsymbol{x}}^{k} \rangle_{\mathrm{s}}
  \right)^{H}
 \right| \boldsymbol{H}^{k}
\right], \\
{\mathcal E}_{\mathrm{SIMO}}^{k,m,\delta} &=  
\mathrm{E}\left[
 \left. 
  \left|
   x_{m}^{k} - \langle \tilde{x}_{m}^{k} \rangle_{\mathrm{s}} 
  \right|^{2}
 \right| \boldsymbol{h}_{m}^{k}
\right], \\
\boldsymbol{\mathcal V}_{\mathrm{MIMO}}^{k,\delta} &=
\mathrm{E}\left[
 \left. 
  \left(
   \tilde{\boldsymbol{x}}^{k} 
   - \langle \tilde{\boldsymbol{x}}^{k} \rangle_{\mathrm{s}}
  \right)
  \left(
   \tilde{\boldsymbol{x}}^{k} 
   - \langle \tilde{\boldsymbol{x}}^{k} \rangle_{\mathrm{s}} 
  \right)^{H}
 \right| \boldsymbol{H}^{k}
\right],  \\
{\mathcal V}_{\mathrm{SIMO}}^{k,m,\delta} &= 
\mathrm{E}\left[ 
 \left. 
  \left|
   \tilde{x}_{m}^{k} - \langle \tilde{x}_{m}^{k} \rangle_{\mathrm{s}}
  \right|^{2}
 \right| \boldsymbol{h}_{m}^{k}
\right], 
\end{align}
\end{subequations}
where the expectations are taken with respect to 
$q(\boldsymbol{y}^{k},\{\boldsymbol{y}_{m}^{k}\}|
\boldsymbol{x}^{k}, \boldsymbol{H}^{k}; \boldsymbol{R})
p(\boldsymbol{x}^{k})$ and  
$\langle \tilde{\boldsymbol{x}}^{k} \rangle_{\mathrm{s}}$ is given by 
\begin{equation} \label{singleuser_posterior_mean_estimator}
\langle \tilde{\boldsymbol{x}}^{k} \rangle_{\mathrm{s}} = 
\frac{
 \mathrm{E}_{\tilde{\boldsymbol{x}}^{k}}\left[ 
  \tilde{\boldsymbol{x}}^{k}
  q(
   \boldsymbol{y}^{k}, \{\boldsymbol{y}_{m}^{k}\} |
    \tilde{\boldsymbol{x}}^{k}, \boldsymbol{H}^{k}; \tilde{\boldsymbol{R}}
  )
 \right]
}
{
 \mathrm{E}_{\tilde{\boldsymbol{x}}^{k}}\left[
  q(
   \boldsymbol{y}^{k}, \{\boldsymbol{y}_{m}^{k}\} |
    \tilde{\boldsymbol{x}}^{k}, \boldsymbol{H}^{k}; \tilde{\boldsymbol{R}}
  )
 \right]
}.
\end{equation}
Since (\ref{singleuser_posterior_mean_estimator}) is 
reduced to (\ref{SIMO_posterior_mean_estimator}) and 
(\ref{MIMO_posterior_mean_estimator}) for $\delta\rightarrow 0$ and 
$\delta\rightarrow 1$, respectively, (\ref{R_0_apen}) and (\ref{R_apen}) 
yield (\ref{R_0_SIMO}) and (\ref{R_SIMO}), and (\ref{R_0_MIMO}) and  
(\ref{R_MIMO}) for $\delta\rightarrow 0$ and 
$\delta\rightarrow 1$, respectively.  

Differentiating the rate function~(\ref{rate_function}) 
with respect to $u$, we obtain 
\begin{equation}
\lim_{u\rightarrow 0}\frac{\partial I^{(u)}}{\partial u}
(\boldsymbol{\mathcal Q}^{\mathrm{s}}) = 
\lim_{K\rightarrow\infty}\frac{1}{K}\sum_{k=1}^{K}
\tilde{\mathcal C}_{\mathrm{s}}^{k}\ln 2
-\frac{1}{\beta}\mathrm{Tr}\left[
 \boldsymbol{I}_{N} - \left(
  N_{0}+\tilde{N}_{0}
 \right)
 \tilde{\boldsymbol{R}}^{-1}
 + \tilde{N}_{0}\tilde{\boldsymbol{R}}^{-1}\boldsymbol{R}
 \tilde{\boldsymbol{R}}^{-1}
\right], \label{deriv_I} 
\end{equation}
where $\tilde{\mathcal C}_{\mathrm{s}}^{k}$ is given by 
\begin{equation}
\tilde{\mathcal C}_{\mathrm{s}}^{k} = 
\mathrm{E}_{\boldsymbol{x}^{k}}\left\{ 
 \int 
  q(
   \boldsymbol{y}^{k},\{\boldsymbol{y}_{m}^{k}\} |
   \boldsymbol{x}^{k}, \boldsymbol{H}^{k}; \boldsymbol{R}
  )
 \log\frac{
 q(
  \boldsymbol{y}^{k},\{\boldsymbol{y}_{m}^{k}\} |
  \boldsymbol{x}^{k}, \boldsymbol{H}^{k}; \tilde{\boldsymbol{R}}
 )
 }
 {
  \mathrm{E}_{\tilde{\boldsymbol{x}}^{k}}\left[
   q(
    \boldsymbol{y}^{k},\{\boldsymbol{y}_{m}^{k}\} |
    \tilde{\boldsymbol{x}}^{k}, \boldsymbol{H}^{k}; \tilde{\boldsymbol{R}}
   )
  \right]
 }
 \boldsymbol{dy}^{k}\prod_{m=1}^{M_{k}}\boldsymbol{dy}_{m}^{k}  
\right\}.
\end{equation} 
On the other hand, (\ref{G}) yields 
\begin{equation}
G^{(u)}(\boldsymbol{\mathcal Q}^{\mathrm{s}}) = 
- (u-1)\ln\det\tilde{\boldsymbol{R}}  
- \ln\det(\tilde{\boldsymbol{R}} + u\boldsymbol{R})
- uN\ln\pi. \label{G_rs}
\end{equation}
Differentiating (\ref{G_rs}) with respect to $u$, we obtain  
\begin{eqnarray}
\lim_{u\rightarrow 0}\frac{\partial G^{(u)}}{\partial u} = 
-\ln\det\tilde{\boldsymbol{R}} 
- \mathrm{Tr}(
 \boldsymbol{R}\tilde{\boldsymbol{R}}^{-1}
) 
- N\ln\pi. \label{deriv_G}
\end{eqnarray}
From (\ref{deriv_I}) and (\ref{deriv_G}),  
the free energy~(\ref{tmp6}) is evaluated as
\begin{eqnarray}
\frac{\beta}{\ln 2}{\mathcal F} = 
\lim_{K\rightarrow\infty}\frac{1}{K}\sum_{k=1}^{K}\beta 
\tilde{\mathcal C}_{\mathrm{s}}^{k}
+ N\log(\pi N_{0}\mathrm{e})
+ F(\boldsymbol{R},\tilde{\boldsymbol{R}}), \label{expect_free_energy_rs} 
\end{eqnarray}
where $F(\boldsymbol{R},\tilde{\boldsymbol{R}})$ is given by 
(\ref{function_F}).  
In particular, (\ref{expect_free_energy_rs}) is reduced to 
(\ref{free_energy_ind}) and (\ref{free_energy_same}) 
for $\delta\rightarrow 0$ and $\delta\rightarrow 1$, respectively. 
In the case in which there exist multiple solutions of (\ref{R_0_apen}) 
and (\ref{R_apen}), one should choose the solution that yields the 
supremum of (\ref{tmp5}), which minimizes the free 
energy~(\ref{expect_free_energy_rs}).      

\section{Derivation of Lemma~\ref{lemma_moment}} 
\label{appendix_joint_moment}
With mutually disjoint subsets $A_{m}^{(\mathrm{r})}$ and 
$A_{m}^{(\mathrm{i})}$ of $\mathcal{A}=\{1,\ldots,u\}$ satisfying  
$|A_{m}^{(\mathrm{r})}|=j_{m}^{(\mathrm{r})}$ and   
$|A_{m}^{(\mathrm{i})}|=j_{m}^{(\mathrm{i})}$, and with nonnegative 
integers $i_{m}^{(\mathrm{r})}$ and $i_{m}^{(\mathrm{i})}$, 
we define a scalar function $f_{k}$ as 
\begin{equation}
f_{k}(\vec{\boldsymbol{\mathcal X}}) = 
\prod_{m=1}^{M_{k}}\left\{
 [\mathrm{Re}(x_{m}^{k,0})]^{i_{m}^{(\mathrm{r})}}
 [\mathrm{Im}({x_{m}^{k,0}})]^{i_{m}^{(\mathrm{i})}}
 \prod_{\alpha_{m}^{\mathrm{(r)}}\in A_{m}^{\mathrm{(r)}}}
 \mathrm{Re}(x_{m}^{k,\alpha_{m}^{\mathrm{(r)}}})
 \prod_{\alpha_{m}^{\mathrm{(i)}}\in A_{m}^{\mathrm{(i)}}}
 \mathrm{Im}(x_{m}^{k,\alpha_{m}^{\mathrm{(i)}}})
\right\}. 
\end{equation}
Furthermore, with a real parameter $\omega$, we define a quantity 
similar to the free energy as 
\begin{equation}
\tilde{{\mathcal F}} = 
\lim_{K,L\rightarrow \infty}\frac{1}{K}\ln\tilde{\Xi}^{(u)}(\omega), 
\label{extended_free_energy} 
\end{equation}
where $\tilde{\Xi}^{(u)}(\omega)$ is given by 
\begin{equation} 
\tilde{\Xi}^{(u)}(\omega) = 
\mathrm{E}\left\{
 \left. 
 \exp\left[
  \omega \sum_{k\in\mathcal{K}_{p}}f_{k}(\vec{\boldsymbol{\mathcal X}})
 \right] 
 \prod_{\alpha=1}^{u}p(\vec{\tilde{\boldsymbol{y}}}=
\vec{\boldsymbol{y}}|\vec{\boldsymbol{x}}^{\alpha}, 
 \boldsymbol{\mathcal S}, \boldsymbol{\mathcal H};\tilde{N}_{0})
 \right| \boldsymbol{\mathcal H}
\right\}.  
\end{equation}
Differentiating (\ref{extended_free_energy}) with respect to $\omega$, 
we obtain 
\begin{equation}
\lim_{u\rightarrow 0}\left.
 \frac{\partial \tilde{\mathcal F}}{\partial \omega}
\right|_{\omega=0} = 
\lim_{u\rightarrow 0}\lim_{K,L\rightarrow \infty}\frac{1}{K}
\sum_{k\in\mathcal{K}_{p}}\mathrm{E}\left\{
 \left.
 f_{k}(\vec{\boldsymbol{\mathcal X}})
 \prod_{\alpha=1}^{u}p(\vec{\tilde{\boldsymbol{y}}}=
\vec{\boldsymbol{y}}|\vec{\boldsymbol{x}}^{\alpha}, 
 \boldsymbol{\mathcal S}, \boldsymbol{\mathcal H};\tilde{N}_{0})
 \right| \boldsymbol{\mathcal H}
\right\}. \label{extended_free_energy_tmp1}
\end{equation}
Substituting the Bayes formula 
\begin{equation} \label{Bayes_formula}
p(\vec{\tilde{\boldsymbol{y}}}|\vec{\boldsymbol{x}}^{\alpha}, 
\boldsymbol{\mathcal S}, \boldsymbol{\mathcal H};\tilde{N}_{0}) = 
\frac{
 p(\vec{\boldsymbol{x}}^{\alpha}|\vec{\tilde{\boldsymbol{y}}}, 
 \boldsymbol{\mathcal S}, \boldsymbol{\mathcal H};\tilde{N}_{0}) 
 \mathrm{E}_{\vec{\tilde{\boldsymbol{x}}}}
  \left[
   p(\vec{\tilde{\boldsymbol{y}}}|
   \vec{\tilde{\boldsymbol{x}}}, \boldsymbol{\mathcal S}, 
   \boldsymbol{\mathcal H};\tilde{N}_{0})
  \right]
}
{
 p(\vec{\boldsymbol{x}}^{\alpha})
}
\end{equation}    
into (\ref{extended_free_energy_tmp1}), we obtain 
\begin{equation}
\lim_{u\rightarrow 0}\left.
 \frac{\partial \tilde{\mathcal F}}{\partial \omega}
\right|_{\omega=0} = 
\lim_{K,L\rightarrow\infty}
\frac{1}{K}\sum_{k\in\mathcal{K}_{p}}
\mathrm{E}\left[
 \left.
  \prod_{m=1}^{M_{k}}\left\{ 
   [\mathrm{Re}(x_{m}^{k})]^{i_{m}^{(\mathrm{r})}}
   [\mathrm{Im}({x_{m}^{k}})]^{i_{m}^{(\mathrm{i})}}
   \langle \mathrm{Re}(\tilde{x}_{m}^{k}) 
   \rangle^{j_{m}^{(\mathrm{r})}}
   \langle \mathrm{Im}(\tilde{x}_{m_{k}}^{k}) 
   \rangle^{j_{m}^{(\mathrm{i})}}
  \right\}
 \right| \boldsymbol{\mathcal H}
\right], \label{joint_PME_moment}
\end{equation}
where we have assumed that the limit with respect to $u$ and 
the many-user limit can be interchanged. 

Next, we evaluate (\ref{extended_free_energy}).  
We obtain the following expression corresponding to (\ref{tmp5}): 
\begin{equation}
\tilde{\mathcal F}
= 
\sup_{\boldsymbol{\mathcal Q}}\left[
 \beta^{-1}G^{(u)}(\boldsymbol{\mathcal Q}) 
 - I^{(u)}(\boldsymbol{\mathcal Q};\omega)
\right], 
\end{equation}
where the extended rate function $I^{(u)}(\boldsymbol{\mathcal Q};\omega)$ 
is given by
\begin{equation}
I^{(u)}(\boldsymbol{\mathcal Q};\omega) = 
\sup_{\tilde{\boldsymbol{\mathcal Q}}}\left[
 \mathrm{Tr}(\tilde{\boldsymbol{\mathcal Q}}\boldsymbol{\mathcal Q}) 
 - \lim_{K\rightarrow\infty}\frac{1}{K}\left( 
  \sum_{k\notin \mathcal{K}_{p}}\ln{\mathcal M}_{k}^{(u)}
  (\tilde{\boldsymbol{\mathcal Q}})
  + \ln\tilde{\mathcal M}^{(u)}(\tilde{\boldsymbol{\mathcal Q}};
  \omega)
 \right)
\right] \label{rate_function_omega}
\end{equation}
with the extended moment generating function of the symbols defined as 
\begin{equation}
\tilde{\mathcal M}^{(u)}(\tilde{\boldsymbol{\mathcal Q}};\omega) 
=
\mathrm{E}\left\{
 \left.
  \exp\left[
  \sum_{k\in\mathcal{K}_{p}}(
   \omega f_{k}(\vec{\boldsymbol{\mathcal X}}) 
    + \Lambda^{k}(\tilde{\boldsymbol{\mathcal Q}}))
  \right]
 \right| \boldsymbol{\mathcal H}
\right\}.
\end{equation}
Differentiating $\tilde{\mathcal F}$ with respect to $\omega$, 
we obtain 
\begin{equation}
\left.
 \lim_{u\rightarrow 0}
 \frac{\partial \tilde{\mathcal F}}{\partial \omega}
\right|_{\omega=0} = 
- \left.
 \lim_{u\rightarrow 0}
 \frac{\partial I^{(u)}}{\partial \omega}
 (\boldsymbol{\mathcal Q}^{\mathrm{s}};\omega) 
\right|_{\omega=0}, \label{extended_free_energy_tmp2}
\end{equation} 
where $\boldsymbol{\mathcal Q}^{\mathrm{s}}$ satisfies  
(\ref{equation1}) and (\ref{equation2}). 

Assuming the replica symmetry, we can evaluate   
(\ref{extended_free_energy_tmp2}) analytically. 
The derivative of the extended moment generating function 
$\tilde{\mathcal M}^{(u)}
(\tilde{\boldsymbol{\mathcal Q}}^{\mathrm{s}};\omega)$ 
with respect to $\omega$ is given by 
\begin{eqnarray}
\left.
 \frac{\partial \tilde{\mathcal M}^{(u)}}{\partial \omega}
 (\tilde{\boldsymbol{\mathcal Q}}^{\mathrm{s}}
 ;\omega) 
\right|_{\omega=0}
=
\sum_{k\in\mathcal{K}_{p}}\int\mathrm{E}_{\boldsymbol{x}^{k}}\left[
 \prod_{m=1}^{M_{k}}\left\{
  [\mathrm{Re}(x_{m}^{k})]^{i_{m}^{(\mathrm{r})}}
  [\mathrm{Im}({x_{m}^{k}})]^{i_{m}^{(\mathrm{i})}}
 \right\}
 q(
  \boldsymbol{y}^{k}, \{\boldsymbol{y}_{m}^{k}\} |
   \boldsymbol{x}^{k}, \boldsymbol{H}^{k}; \boldsymbol{R}
 )
\right] \nonumber \\ 
\prod_{m=1}^{M_{k}}\left\{
 \left\langle 
  \mathrm{Re}(\tilde{x}_{m}^{k}) 
 \right\rangle_{\mathrm{s}}^{j_{m}^{(\mathrm{r})}}
 \left\langle 
  \mathrm{Im}(\tilde{x}_{m_{k}}^{k}) 
 \right\rangle_{\mathrm{s}}^{j_{m}^{(\mathrm{i})}} 
\right\}
\left\{
 \frac{
  \mathrm{E}_{\tilde{\boldsymbol{x}}^{k}}\left[
   q(
    \boldsymbol{y}^{k}, \{\boldsymbol{y}_{m}^{k}\} |
     \tilde{\boldsymbol{x}}^{k}, \boldsymbol{H}^{k}; \tilde{\boldsymbol{R}}
   ) 
  \right]
 }
 {
 q(
  \boldsymbol{y}^{k}, \{\boldsymbol{y}_{m}^{k}\} |
   \boldsymbol{0}, \boldsymbol{H}^{k}; \tilde{\boldsymbol{R}}
 ) 
 }
\right\}^{u}
\boldsymbol{dy}^{k}\prod_{m=1}^{M_{k}}\boldsymbol{dy}_{m}^{k}, 
  \label{moment_generating_omega_deriv}
\end{eqnarray} 
where $\boldsymbol{R}$ and $\tilde{\boldsymbol{R}}$ satisfy 
(\ref{R_0_apen}) and (\ref{R_apen}). 
From (\ref{joint_PME_moment}), (\ref{rate_function_omega}), 
(\ref{extended_free_energy_tmp2}), and (\ref{moment_generating_omega_deriv}),  
we obtain  
\begin{eqnarray}
\lim_{K,L\rightarrow\infty}
\frac{1}{|\mathcal{K}_{p}|}\sum_{k\in\mathcal{K}_{p}}
\mathrm{E}\left[
 \left.
  \prod_{m=1}^{M_{k}}\left\{
   [\mathrm{Re}(x_{m}^{k})]^{i_{m}^{(\mathrm{r})}}
   [\mathrm{Im}({x_{m}^{k}})]^{i_{m}^{(\mathrm{i})}}
   \langle \mathrm{Re}(\tilde{x}_{m}^{k}) 
   \rangle^{j_{m}^{(\mathrm{r})}}
   \langle \mathrm{Im}(\tilde{x}_{m_{k}}^{k}) 
   \rangle^{j_{m}^{(\mathrm{i})}}
  \right\} 
 \right| \boldsymbol{\mathcal H}
\right] \nonumber \\ 
= 
\lim_{|\mathcal{K}_{p}|\rightarrow \infty}\frac{1}{|\mathcal{K}_{p}|}
\sum_{k\in\mathcal{K}_{p}}\mathrm{E}\left[
 \left. 
  \prod_{m=1}^{M_{k}}\left\{ 
   [\mathrm{Re}(x_{m}^{k})]^{i_{m}^{(\mathrm{r})}}
   [\mathrm{Im}({x_{m}^{k}})]^{i_{m}^{(\mathrm{i})}}
   \left\langle 
    \mathrm{Re}(\tilde{x}_{m}^{k}) 
   \right\rangle_{\mathrm{s}}^{j_{m}^{(\mathrm{r})}}
   \left\langle 
    \mathrm{Im}(\tilde{x}_{m}^{k}) 
   \right\rangle_{\mathrm{s}}^{j_{m}^{(\mathrm{i})}}
  \right\}
 \right| \boldsymbol{H}^{k}
\right].  
\end{eqnarray} 


\section*{Acknowledgment}
TT acknowledges support through Grant-in-Aid for Scientific 
Research on Priority Areas (No. 18079010) from MEXT, Japan.

\ifCLASSOPTIONcaptionsoff
  \newpage
\fi



%
%
%

\bibliographystyle{IEEEtran}
\bibliography{IEEEabrv,mimo_cdma}

\newpage 

\newpage

\newpage 

\newpage

%








\end{document}